\documentclass[pdflatex,sn-mathphys-num,iicol]{sn-jnl}

\usepackage{arydshln}
\usepackage{stmaryrd}
\usepackage{graphicx}%
\usepackage{multirow}%
\usepackage{amsmath,amssymb,amsfonts}%
\usepackage{amsthm}%
\usepackage{mathrsfs}%
\usepackage[title]{appendix}%
\usepackage{xcolor}%
\usepackage{textcomp}%
\usepackage{manyfoot}%
\usepackage{booktabs}%
\usepackage{algorithm}%
\usepackage{algorithmicx}%
\usepackage{algpseudocode}%
\usepackage{listings}%
\usepackage[T1]{fontenc}
\usepackage{ dsfont }
\usepackage{mathtools}
\usepackage{subcaption}
\usepackage{anyfontsize}

\newtheorem{remark}{Remark}%

\begin{document}

\title[Article Title]{Calibration of stress--jump conditions for arbitrary flow directions in fluid--porous systems}


\author[1]{\sur{Philippe Angot}}\email{philippe.angot@univ-amu.fr}

\author*[1]{\sur{Joscha Nickl}}\email{joscha.nickl@univ-amu.fr}


\affil*[1]{\orgdiv{Institut de Mathématiques de Marseille}, \orgname{Aix-Marseille Université }, \orgaddress{\street{3 pl. V. Hugo - Case 19}, \city{Marseille Cedex 03}, \postcode{13331},  \country{France}}}




\newcommand{\FF}{{\mathrm{f}}}
\newcommand{\PM}{{\mathrm{p}}}
\newcommand{\todo}[1]{{\color{red!80!black} {ToDo: #1}}}
\newcommand{\vdot}{\boldsymbol{\mathsf{\ensuremath\cdot}}}

\newcommand{\pd}[2]{\frac{\partial{#1}}{\partial{#2}}}
\newcommand{\pdd}[2]{\frac{\partial^2{#1}}{\partial{#2}^2}}
\newcommand{\del}{\ensuremath{\nabla}}
\newcommand{\deld}{\ensuremath{\del\vdot}}%
\newcommand{\lrp}[1]{\left( #1 \right)}
\newcommand{\norm}[2]{\left\| #1 \right\|_{#2}}
\renewcommand{\vec}[1]{{\ensuremath{\boldsymbol{\mathrm #1}}}}
\newcommand{\BJ}{{\mathrm{BJ}}}
\newcommand{\dsp}{\displaystyle}

\abstract{

A numerical validation of the stress‑jump coupling conditions for Stokes–Darcy flow in two dimensions is presented, addressing a gap that has remained since their introduction by Angot et al. \citep{Angot_etal_17}. These conditions, formulated for arbitrary flow directions at the interface between a porous medium and an adjacent free‑flow region, involve a friction tensor whose coefficients are not known a priori. We calibrate these parameters for a range of porous‑medium configurations and flow regimes by matching the macroscopic model to reference solutions derived from processed pore‑scale simulations. Several optimization strategies are assessed for this calibration task. The results show that, although three parameters are formally required, exploiting structural properties of the porous medium enables an effective reduction to a one‑dimensional calibration with negligible loss in accuracy. A regional sensitivity analysis further indicates that even coarse parameter estimates can yield a well‑performing model, highlighting the robustness and practical applicability of the stress‑jump formulation.

}




\maketitle

\section{Introduction}
\label{sec:introduction}

Coupled flows refer to the interactions between the flow in a porous medium and flow in an adjacent free-fluid domain. This phenomenon is critical in applications such as groundwater management, biological systems, and various industrial processes \citep{DAS2002,DANGELO2011,DEREIMS2015}.

The flow can be considered on different spatial scales. On the microscale or porescale, the pores are resolved. In practice this can be done for instance using computed tomography (CT) scans of a porous medium \citep{narsilioUpscalingNavierStokes2009,fredrichPredictingMacroscopicTransport2006}. The mathematical models to describe  flow at microscale  are classical equations of fluid dynamics like the Stokes or Navier--Stokes equations. However, such simulations are computationally costly and thus restricted to small domains because the geometry of the pores has to be resolved. Additionally, for larger flow problems, it is often impossible to determine the geometry of the porous medium. To avoid this problem, coupled flow can be considered instead at the macroscale.
In the porous medium for this purpose, 
the equations can be derived from pore-scale models through techniques like two-scale homogenization or volume averaging. In this way, one can obtain for instance Darcy's law from the Stokes equations in a porous medium \citep{Hornung1997,WHITAKER1986a}. This upscaling process results in a macroscale model that does not require a detailed resolution of the pore structure, but instead uses effective parameters such as the permeability tensor. These parameters can be determined experimentally, using correlations connecting properties of the porous medium on pore-scale as the porosity and effective parameters \citep{ELHAKIM2016,VERVOORT2003} or by performing simulations on pore-scale. The typical examples of upscaled equations are Darcy's law, Brinkman's and Forchheimer's equations.

In order to solve coupled-flow problems, where in addition to the porous medium a free-flow region is present, different approaches can be followed. To consider coupled flows at the macroscale, there exist the one-domain approach and the two-domain approach. 
In the one-domain approach, the same system of partial differential equations (PDE) is considered in the region of the free flow and porous medium. The change from free-flow to porous medium is modeled by variation of parameters in the problem formulation \citep{VALDES-PARADA2021,HERNANDEZ-RODRIGUEZ2022}.

Apart from direct numerical simulations, we will concentrate in this work on the two-domain approach. This approach involves using two distinct sets of partial differential equations: one for the free-flow and another one for the porous medium, coupled by interface conditions at their common boundary. The free flow is governed by classical fluid dynamics equations. In the porous medium upscaled equations are employed. The choice of interface conditions, the parameters appearing therein and the position of the interface are the main modelling aspects for this approach. 
On the interface between the free flow and porous medium, typically coupling conditions  are written to ensure mass conservation. Furthermore, either the balance of normal forces, or the continuity of the pressure are used as discussed in \citep{LYU2021}. In \citep{CARRARO2013}, a version of the continuity of pressure was considered. The well-known Beavers--Joseph condition \citep{BEAVERS1967}, extended by Jones \citep{JONES1973}, addresses a flow parallel to the interface  and accommodate a jump in the tangential velocity component. A historical review on the Beavers--Joseph condition can be found in \citet{NIELD2009}. These conditions exhibit the so-called Beavers--Joseph slip parameter $\alpha_\BJ$, which has to be determined in a calibration \citep{YANG2019,STROHBECK2023,MIERZWICZAK2019}.
Nevertheless, the traditional conditions may exhibit reduced accuracy when applied to the case of arbitrary flow directions \cite{EGGENWEILER2020}. Additionally, in the context of inertial flow, the Beavers--Joseph parameter becomes spatially variable \cite{YANG2019}. These observations motivate the formulation of alternative interface conditions. The validation of interface conditions often takes place by comparison of the macroscale model with a model on pore scale.

Whereas some conditions are similar to the classical ones, stress--jump conditions for the tangential stress have been previously proposed by \citet{OCHOA-TAPIA1995,OCHOA-TAPIA1995a} and \citet{VALDES-PARADA2013} for one-dimensional flow 
when the flow in the porous medium is governed by Brinkman's equations. In this article, we will consider the stress--jump conditions developed in \citet{Angot_etal_17} by means of asymptotic analysis. The key idea of its derivation is to consider between free flow and porous medium a highly porous layer. For this situation, a physically reasonable macroscale model is proposed. The coupling conditions are obtained by integrating the governing equations in the  transition zone with respect to the height of the zone. By choosing different quadrature rules for approximating the resulting integrals, different sets of interface conditions can be obtained by dimensional reduction. In the stress--jump conditions, the friction tensor that corresponds to the average of the permeability in the transition zone appears. The value of this tensor is a priori unknown.  In \citet{Angot_etal_21} a calibration of the stress--jump has been performed for one-dimensional flows using the solution of different macroscale problems as a reference. The coupled Stokes/Brinkman and Stokes/Darcy systems are proven to be well-posed for symmetric positive semi-definite friction tensors \cite{Angot2021}.

Many mathematical models for complex applications involve parameters that are initially unknown. In order to effectively use such a model, these parameters have to be determined. Finding these parameters "by hand",  i.e., trying different values and examining the simulation results can be tedious and limits strongly the amount of tested parameter configurations. A remedy to such problems is an automatic calibration. Automatic calibration approaches transform the parameter determination process into an optimization problem \citep{HASSAN2022}. The calibration process begins with defining a cost function within a search space encompassed by the parameter space. This cost function typically quantifies the simulation results against a known ground truth. Simulated annealing and versions thereof are a popular choice for such a procedure, as they can surpass local minima and proved useful in many applications \citep{MATOTT2011}.

The objective of this study is to carry out a thorough numerical calibration of the Stokes--Darcy system with stress--jump conditions proposed in \citep{Angot_etal_17}.  Various test cases and porous--medium configurations are considered. We consider test cases with inflow as well as with outflow of the porous medium at different angles. We compare different search strategies for the parameter search and perform a regional sensitivity analysis. This gives insights on how to efficiently choose parameters for the stress--jump conditions, reducing the complexity of the calibration problem. 

This paper is organized as follows: Section~\ref{sec:models} presents the mathematical models, including the coupled models and their finite element implementation. Section~\ref{chap:Calibration} details the calibration procedure, including the generation of reference solutions and optimization strategies. Afterwards the regional sensitivity analysis approach is presented. In Section~\ref{chap:numericalexperiments} we discuss the numerical experiments conducted on three different geometries and compare the results with existing calibration methods. Finally, Section~\ref{chap:conclusion} summarizes the findings and their implications.

\section{Mathematical models}\label{sec:models}
In this section, we introduce the partial differential equations that describe the coupled flow at both the pore and macroscale. We focus on the stress--jump condition but also consider the classical Beavers--Joseph--Jones interface conditions. Afterwards, we discuss the numerical schemes to solve these equations.
\subsection{Pore-scale and macroscale models}

\begin{figure*}[t]
    \centering
    \includegraphics[width=0.49\textwidth]{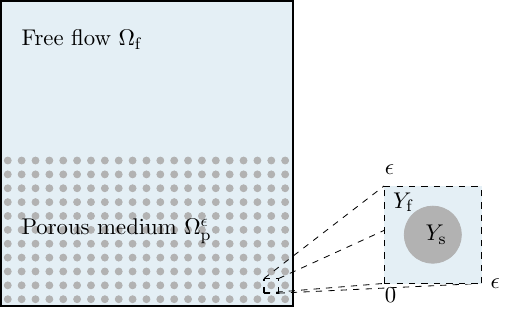}
    \quad\quad
    \includegraphics[width=0.29\textwidth]{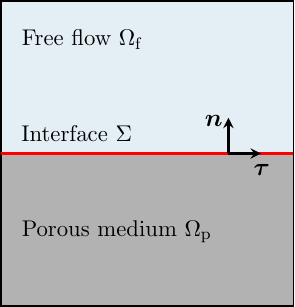}
    \caption{Geometry on pore-scale with scaled unit cell (left) and geometry on macroscale (right)}
    \label{fig:geometries}
\end{figure*}
We consider polygonal domains in two-dimensional space. At the pore scale, the porous medium denoted as $\Omega^\epsilon_\PM\subset\mathbb{R}^2$ consists of periodically distributed impermeable rigid inclusions. The parameter $\epsilon$ denotes the periodicity, corresponding to the length of the scaled unit cell, as illustrated  in Fig.~\ref{fig:geometries} (left). The free-flow domain $\Omega^\FF$ is separated from the porous medium by an interface $\Sigma\subset\mathbb{R}^1$. We define the union of these domains as $\Omega^\epsilon$$\coloneqq \Omega_\FF\cup\Omega_{p}^\epsilon\cup\Sigma$.
We will consider a one-phase, one-species, incompressible, isothermal, steady flow at low Reynolds numbers ($Re\ll 1$). The velocity field of a flow will be denoted throughout this article by the letter $\vec v$, and the corresponding pressure by $p$ with additional sub- and superscripts indicating the mathematical problem they belong to.
This flow is governed by the stationary Stokes equations
    \begin{equation}
\begin{split}
 \begin{alignedat}{2}\label{eq:poreStokes}
    -\deld\mathbf{T}^\FF\lrp{\vec v^\epsilon,p^\epsilon} &= 
    \vec 0
    \, \quad &&\text{in } \ \Omega^\epsilon  \,  , 
    \\
    \nabla \vdot \vec v^\epsilon &= 0 \quad  &&\text{in } \ \Omega^\epsilon  ,
    \end{alignedat}
\end{split}
\end{equation}
where the stress tensor reads 
\begin{align*}
    \mathbf{T}^\FF\lrp{\vec v^\epsilon,p^\epsilon} &= \mu\lrp{\del \vec v^\epsilon+\lrp{\del \vec v^\epsilon}^\top}-p\mathbf{I}\,.
\end{align*}
This system is supplemented by appropriate boundary conditions on $\partial\Omega^\epsilon$. 
The dynamic viscosity of the fluid is denoted by $\mu$.

A macroscale domain is shown on the right hand side of Fig.~\ref{fig:geometries}.
 At the macroscale, the porous--medium region is represented by $\Omega_\PM$ and we choose the normal vector $\vec n$ at the fluid--porous interface to be directed away from the porous medium. 
 
In the free-flow domain, we consider the Stokes equations
\begin{equation}
\begin{split}
 \begin{alignedat}{2}\label{eq:Stokesff}
     -\deld\mathbf{T}^\FF\lrp{\vec v^\FF,p^\FF} &= 
    \vec 0
    \, \quad &&\text{in } \ \Omega_\FF \,  , 
    \\
    \nabla \vdot \vec v^\FF &= 0 \quad  &&\text{in } \ \Omega_\FF \,  .
\end{alignedat}
\end{split}
\end{equation}
In the porous medium, Darcy's flow equations 
\begin{equation}
\begin{split}
 \begin{alignedat}{2}\label{eq:Darcypm}
         \vec v^\PM &= - \frac{\mathbf{K}}{\mu}\,\nabla p^\PM  = \frac{\mathbf{K}}{\mu}\,\nabla \vdot \mathbf{T}^\PM
 \, \quad  &&\text{in } \ \Omega_\PM \,  , 
\\
\nabla \vdot \vec v^\PM &= 0 \quad 
&&\text{in } \ \Omega_\PM \,  ,
\end{alignedat}
\end{split}
\end{equation}\\
are used to describe the flow. The stress tensor in the porous medium is $\mathbf{T}^p(p^\PM)=-p^\PM\mathbf{I}$. The permeability tensor $\mathbf{K}=\{K_{i j}\}_{i,j=1,2}$ is symmetric positive definite. In addition, suitable boundary conditions must be imposed on $\partial\Omega_p\setminus \Sigma$. Unlike the Stokes equations, Darcy's law describes the velocity field in terms of an average velocity, rather than the velocity of the fluid at a specific location in space. For periodic porous media, the entries of the permeability tensor $\mathbf{K}$ can be determined solving auxiliary problems. These auxiliary problems consist of the Stokes equations defined in a unit cell, as illustrated in Fig.~\ref{fig:geometries} in its scaled version. According to homogenization theory \citep{Hornung1997}, we have
\begin{subequations}\label{cellproblem}
\begin{alignat*}{2}
     -\deld \lrp{\nabla \vec w^j\!+\lrp{\del  \vec w^j}^\top}+\nabla \pi^j&=\vec e_j&& \text{ in } Y_f\,,\phantom{aaaaaa}\\
    \deld \vec w^j&=0&&\text{ in }Y_\FF\,,\\
   \vec w^j&=\vec 0 &&\text{ on }\partial Y_\text{s}\,,\\
\end{alignat*}
    \end{subequations}
    where $\vec w^j,\pi^j$ are 1-periodic and $j\in \{1,2\}\,$. 

By averaging and scaling the solutions with the periodicity parameter $\epsilon$ (see Fig.~\ref{fig:geometries}, left), we obtain the formulation for the permeability
    \begin{align}\label{def:permeability}
    K_{ij}=\epsilon^2\int_{Y_f} w^{j}_i\lrp{\vec y}\mathrm{d}\vec y\,.\end{align}
On the interface $\Sigma$, the Stokes and Darcy equations need to be coupled to achieve an exchange of information and a closed model formulation.

The most widely used conditions to couple the two systems \eqref{eq:Stokesff},\eqref{eq:Darcypm} are the interface conditions below:
\begin{subequations}\label{eq:classicalic}
    \begin{eqnarray}
            \llbracket     \vec v\vdot\vec n\rrbracket_\Sigma=& 0\phantom{aaaaaaaaa}&\ \ \text{on } \ \Sigma  \, ,\label{eq:COMclass}\\
   \llbracket \mathbf{T}({\vec v,p}) \vec n\rrbracket_\Sigma\vdot\vec n=&0\phantom{aaaaaaaaa}&\ \ \text{on } \ \Sigma \, ,\label{eq:BONF}\\
   \llbracket     \vec v\vdot \vec\tau\rrbracket_\Sigma=&\dsp\frac{\alpha_\BJ}{\sqrt{\mathbf{K}}}\mathbf{T}(\vec v^\FF,p^\FF)\vec n\vdot \vec \tau &\ \ \text{on } \ \Sigma \,. \quad\label{eq:BJ}
    \end{eqnarray}
\end{subequations}
The normal vector $\vec n$ and the tangential vector $\vec \tau$ at the interface are depicted in Fig.~\ref{fig:geometries}.
For a quantity $\phi$ defined in $\Omega_\FF$ by $\phi^\FF$ and in $\Omega_\PM$ by $\phi^\PM$ we use the notation\begin{align}
    \llbracket \phi\rrbracket_\Sigma = \phi^\FF-\phi^\PM\,.
\end{align}
The first condition \eqref{eq:COMclass} ensures, that on the interface the conservation of mass holds. This implies that no mass is stored or created on the interface. The second condition \eqref{eq:BONF} states that on the interface the normal forces from the porous medium and free flow are balanced. The third condition is the Beavers--Joseph--Jones condition that was in its original form first proposed by \citet{BEAVERS1967} as an ad hoc condition. It was later extended by \citet{JONES1973} for symmetric viscous stress. The parameter $\alpha_\BJ$ appearing in these conditions is initially unknown and requires a calibration \citep{YANG2019,STROHBECK2023,MIERZWICZAK2019}. It is a positive parameter. For the factor $\sqrt{\mathbf{K}}$ different definitions exist \citep{EGGENWEILER2020}, that in the case of isotropic media coincide. 

We choose $\sqrt{\mathbf{K}}=\sqrt{(k_{11}+k_{22})/2}$. 
We will denote conditions \eqref{eq:classicalic} in the following as classical. 
For the well-posedness of the equations \eqref{eq:Stokesff},\eqref{eq:Darcypm},\eqref{eq:classicalic} supplemented by boundary conditions, we refer to \citep{Cao_Gunzburger_etal_10} and \citep{Angot_2011}.

A different set of interface conditions are stress--jump conditions. They were first developed for the coupling of the Stokes and Brinkman equations in~\cite{OCHOA-TAPIA1995a, VALDES-PARADA2013} before being derived for both the Stokes--Brinkman and Stokes--Darcy problems in \citet{Angot_etal_17}. For the Stokes--Darcy coupling the conditions consist of the conservation of mass
\begin{subequations}\label{eq:SJic}
\begin{equation}
              \llbracket     \vec v\rrbracket_\Sigma\vdot\vec n=0\phantom{aaaaaaaaa}\ \ \text{on } \ \Sigma  \, ,
\end{equation}
as it is used in \eqref{eq:classicalic}. Additionally, a condition on the  jump of the stress vectors
\begin{equation}
    \llbracket \mathbf{T}({\vec v,p})\vec n\rrbracket_\Sigma=\dsp\frac{\mu}{\sqrt{\mathbf{K}}}\vec\beta\vec v^\FF\ \ \text{on } \ \Sigma \,, \label{eq:SJ}
\end{equation}
\end{subequations}
is imposed.
The friction tensor $\vec\beta$ is an unknown quantity. Thus, the stress--jump model needs to be calibrated. Due to the tensorial form of the friction tensor and the general situation considered for its derivation, these conditions are supposed to handle flow with arbitrary directions to the interface well. The friction tensor is supposed to be symmetric positive semidefinite. From now on, we will refer to these conditions as the stress--jump conditions \eqref{eq:SJic}. For one-dimensional flows, a calibration of the stress--jump conditions was carried out by \citet{Angot_etal_21}. By comparing the stress--jump problems to a one-domain macroscale model a scalar friction tensor was determined. Furthermore this study found as an approximation for a scalar friction tensor
\begin{equation}\label{eq:IC-beta_1d-SJ}
\displaystyle\vec\beta \approx \frac{1}{\sqrt{\mathrm{Da}}},
\end{equation}
with the Darcy number $\mathrm{Da} := \frac{K}{L^2}$ being defined in terms of a characteristic macroscopic length $L$ and the scalar-valued permeability $K$ (for isotropic porous media).

 The well-posedness analysis of the Stokes--Darcy system with the stress--jump conditions \eqref{eq:Stokesff}, \eqref{eq:Darcypm}, \eqref{eq:SJic} and suitably chosen boundary conditions is performed in \citet{Angot2021}.

 We will  mainly consider \eqref{eq:Stokesff}, \eqref{eq:Darcypm}, \eqref{eq:SJic} with calibrated friction tensor $\vec\beta$. We will compare the calibrated model with a similarly calibrated classical model \eqref{eq:Stokesff}, \eqref{eq:Darcypm}, \eqref{eq:classicalic}.

\subsection{Finite element implementation of pore- and macroscale problems}\label{subsec:FEMimplementation}


We use a finite element method to solve the coupled Stokes--Darcy problems as well as the Stokes problem on the pore scale, numerically. 
The discrete problem is formulated in a monolithic manner.

We denote by $T^\FF$ and $T^\PM$ the triangulation of the free-flow  and porous-medium domain, respectively. On the interface,  $T^\FF$ and  $T^\PM$ share the same nodes and edges.
The finite dimensional spaces constructed on these meshes using the Taylor--Hood elements are denoted by $V^\FF$ for the free flow and $V^\PM$ for the porous medium. 
In order to define the global stiffness matrix of the problem we identify several sub-problems that together form the Stokes--Darcy problem \eqref{eq:Stokesff}, \eqref{eq:Darcypm} and \eqref{eq:SJic}. For better readability we formulate the sub-problems as linear and bilinear forms. We neglect contributions that stem from boundary conditions (apart from the interface conditions) for the corresponding problems.  
The bilinear form 
\begin{equation}
\begin{split}
 \begin{alignedat}{2}\label{eq:bilinearformstokes}
    S(\{\vec u,p\}&,\{\vec v,q\})\\
    =& \mu\int_{\Omega_\FF}\lrp{\del \vec u+\lrp{\del \vec u}}\mathbf{:}\del\vec v
    \\
    &-\int_{\Omega_\FF} p\deld \vec v+\int_{\Omega_\FF} q\deld \vec u   -\int_{\Omega_\FF}\varepsilon_p p q\\
    &+\int_\Sigma\frac{\mu}{\sqrt{\mathbf{K}}}  \lrp{\vec\beta \vec u}\vdot \vec v\,
\end{alignedat}
\end{split}
\end{equation}defined on the space $V^\FF\times V^\FF$ corresponds to the discretisation discussed in chapter 5.6 of the documentation of FreeFem++ \cite{Hecht2012} in its release version 4.8.  We choose the penalization parameter  $\varepsilon_p=10^{-12}$. The surface integral over the interface $\Sigma$ is a part of the encoding of the stress--jump conditions \eqref{eq:SJic}. 

In \citet{Correa2008}, a stabilized mixed formulation for the Darcy problem is introduced. For the Taylor--Hood elements this scheme was adapted to coupled Stokes--Darcy problems \citep{Correa2009}. The scheme was originally proposed for isotropic porous media. We extend this scheme to the case of general symmetric permeability tensors. Namely, the corresponding bilinear form on $V^\PM\times V^\PM$  below
\begin{align}\label{eq:bilinearformdarcy}
    D(\{\vec u,p\},&\{\vec v,q\})\notag =\int_{\Omega_\PM} \mathbf{K}^{-1}\vec u\vdot \vec v \\
    &-\int_{\Omega_\PM}p\deld \vec v-\int_{\Omega_\PM}q\deld \vec u\notag\\
    &+\int_{\Omega_\PM}\lrp{\deld\vec u}\lrp{ \deld\vec v}\\
    &+s\int_{\Omega_\PM} \operatorname{curl}\lrp{\mathbf{K}^{-1}\vec u}\operatorname{curl}\lrp{\mathbf{K}^{-1}\vec v}\notag
\end{align}
is used. We use the 2D differential operator $\operatorname{curl}(u_1,u_2)= \partial_x u_2-\partial_x u_1$ on the corresponding domain and the rescaling
\begin{align*}
    s=([K_{11}|+|K_{22}|+|K_{12}|+|K_{12}|)\,.
\end{align*}

The implementation of the stress--jump interface conditions is formulated using the bilinear forms on $V^\PM\times V^\FF$:
\begin{align}\label{eq:bilinearformsjmix}
    SJ(\{\vec u,p\},\{\vec v,q\})=\int_\Sigma p \vec{n} \vdot\vec v\,.
\end{align}
The stiffness matrices  $S$, $D$, $SJ$ corresponding  to the bilinear forms \eqref{eq:bilinearformstokes}, \eqref{eq:bilinearformdarcy}, \eqref{eq:bilinearformsjmix}  are created respectively. Our aim is to build a block matrix $SD$ corresponding to the coupled problem. So far we are able to assemble the matrix
\begin{align*}
 \widehat{SD}=\left(
\begin{array}{c:cc}
S & SJ \\ \hdashline
0 & D 
\end{array}
\right)\,.
\end{align*}
The continuity of the normal velocity \eqref{eq:COMclass} can be implemented by penalization.
Let $i$ denote the index of the node for  $\mathcal{T}^\PM $ and $j$ the index of the nodes in  $\mathcal{T}^\FF$, then we define the matrices IntD $\in\mathbb{R}^{\#dof D\times \#dof D}$ with $\#dof D$ the number of degrees of freedom for the porous medium discretizations
  \begin{align*}
  (\text{IntD})_{i,i} = 
  \begin{cases}
  10^{7} & \text{if node for $ \vec v^\PM\vdot \vec n$ } i \text{ on } \Sigma \\
  0 & \text{else } 
  \end{cases}
  \end{align*}
  and IntS $\in\mathbb{R}^{\#dof D\times \#dof S}$ with $\#dof S$ the number of degrees of freedom for the free-flow medium discretizations 
  \begin{align*}
  (\text{IntS})_{i,j} = 
  \begin{cases}
   10^{7} & \text{if node for $ \vec v^\PM\vdot \vec n$ } i \text{ on } \Sigma \\
     & \text{\& if node for $ \vec v^\FF\vdot \vec n$ } j \text{ on } \Sigma \\
 \\
  0 & \text{else}\,.
  \end{cases}
  \end{align*}
We can then assemble the stiffness matrix for the coupled problem
\begin{align*}
 SD=\left(
\begin{array}{c:cc}
S & SJ \\ \hdashline
\text{IntS} & D-\text{IntD} 
\end{array}
\right)
\end{align*}

A similar implementation consists of erasing the rows that correspond to the degrees of freedom ($dof$) on the interface in $D$ for the velocity component normal to the interface. At their place we insert in the block matrix rows that encode the continuity of the normal velocity \eqref{eq:COMclass} on the interface. So, we end up with

\begin{align}
 \widetilde{SD}=\left(
\begin{array}{c:cc}
S & SJ \\ \hdashline\\[-1em]
\widetilde{C} & \widetilde{ D}\phantom{ I} 
\end{array}
\right)
\end{align}

Both versions of the implementation of the continuity of normal velocities \eqref{eq:COMclass} lead to the same experimental order of convergence.

The corresponding code consists of Freefem++ and Python parts that are related by PyFreeFEM \cite{Feppon2020}.

We performed an experimental convergence analysis of the discretization approach for an exact solution based on the Taylor--Green vertices, given in Appendix \ref{app:EOC}. The results are depicted in Fig.~\ref{fig:EOCtest}.

\begin{figure}
  \centering
  \subfloat[quantities in free-flow]{\includegraphics[width=0.4\textwidth]{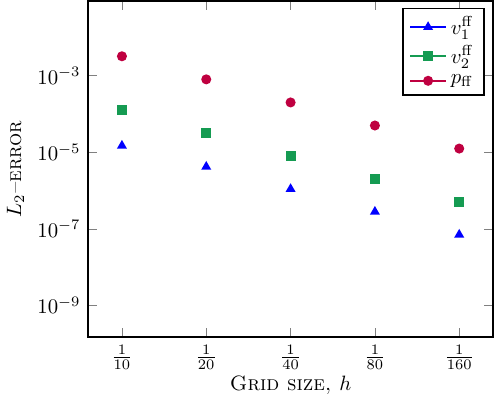}}
  \hfill
  \subfloat[quantities in porous medium flow]{\includegraphics[width=0.4\textwidth]{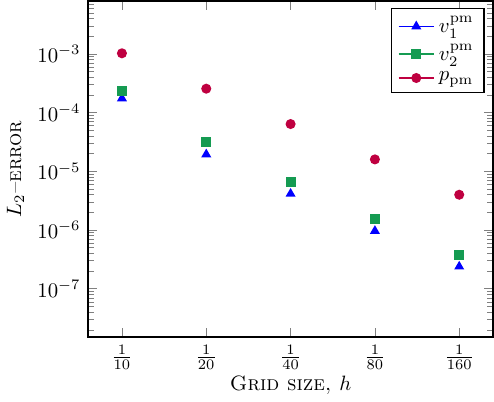}}
  \caption{Convergence analysis for  velocity and pressure \label{fig:EOCtest}}
\end{figure}

\section{General procedure of calibration}\label{chap:Calibration}
In this section, we describe the fundamental aspects of the calibration technique that we employ. The term calibration refers to determining the friction tensor $\vec \beta$ that ensures the macroscale simulations agree with the pore‑scale simulations. At the heart of the process lies the comparison of macroscale simulations for the Stokes--Darcy problem with the stress--jump conditions \eqref{eq:SJic} with a suitable reference solution. The creation of this reference solution is detailed in the next sub-section. Afterwards the optimization with respect to the friction tensor $\vec\beta$ is discussed. We first introduce the search spaces considered before turning to the optimization algorithms that are applied.

\subsection{Comparison of simulation results}
In this section, we precise how the reference solution is obtained from the pore-scale simulations. Afterwards we define the cost function used to quantify the mismatch of the reference and macroscale solutions.

\subsubsection{Manufacturing of reference solutions}\label{subsec:ManufacturingOfReferenceSolutions}
In order to determine how well a coupled system with a fixed friction tensor captures the characteristics of a flow, we need to find a reference solution that we can consider as ground truth. In \citet{Angot_etal_21} the reference solution was chosen as the solution to another macroscale problem -- specifically, a single-domain model. In contrast to this approach, we mimic the derivation of Darcy's law from the pore-scale problem by volume averaging \citep{VALDES-PARADA2021}. We start by solving the stationary Stokes equations in $\dsp \Omega^\epsilon$. 

From the theory of volume averaging it is known that the superficial average for the velocity
\begin{align*}
    \vec v^{avg}(\vec a)=\frac{1}{|V(\vec a)|}\int_{V_f(\vec a)} \vec v(\vec x)\mathrm{d}\vec x\,
\end{align*}
and intrinsic pressure 
\begin{align*}
    p^{avg}(\vec a)=\frac{1}{|V_f(\vec a)|}\int_{V_f(\vec a)}  p(\vec x)\mathrm{d}\vec x\,
\end{align*}
in a porous medium approximate the solutions of the Darcy equations at the point $\dsp\vec a=(a_1,a_2)^\top$, for a suitable representative elementary volume (REV) $V(\vec a)$ \citep{WHITAKER1986a}. The representative elementary volume for  the point $\vec a$ denoted by $V(\vec a)$, is a volume inside the porous medium $\Omega_\PM$ entailing the point $\vec a$. Its dimensions are usually chosen between characteristic length scales for the micro- and macroscale size \citep{WHITAKER1986a}. The fluid phase of the REV is denoted by $V_f(\vec a)$. 

\begin{figure}
    \centering
    \includegraphics[width=0.95\linewidth]{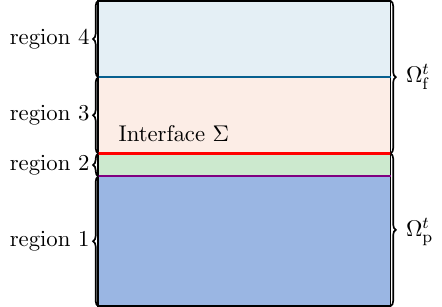}
    \caption{Manufacturing of reference solution}
    \label{fig:referencesol}
\end{figure}

We divide the simulation domain in four regions, that are treated differently as it is depicted in Fig.~\ref{fig:referencesol}.
 By choosing the REV according to the test case and region differently, we can improve the resemblance of this ground truth to the solution of a simulation result stemming from a two-domain approach.
Since an average of a quantity can only be defined as long as the REV lies in the domain of definition, we introduce the domains 
\begin{align*}
    \Omega_\PM^t\coloneqq\{&\vec x=(x,y)^\top|\vec x\in\Omega_\PM, \\&
    \operatorname{dist}(\vec x,\partial (\Omega_\PM\cup\Omega_\FF\cup\Sigma)\geq\frac{\epsilon}{2} \}\,,\\
    \Omega_\FF^t\coloneqq\{&\vec x=(x,y)^\top|\vec x\in\Omega_\FF, \\& \operatorname{dist}(\vec x,\partial (\Omega_\PM\cup\Omega_\FF\cup\Sigma)\geq\frac{\epsilon}{2}  \text{ and }y>0.5 \}\,.
\end{align*}
where the reference solution $\vec v^r$, $p^r$ can be defined in the following.

 In the regions 1, 2 and 3 we use volume averaging with different sizes of the REV. In the forth region, we use directly the pore-scale solution.

The first region is the interior of the porous medium that contains all points in the porous medium with a distance from the interface larger than $\dsp\epsilon/2$.

The second region is the zone inside the porous medium that is closer than $\dsp \epsilon/2$ to the interface. Close to the interface, the correct choice of averaging is not clear \citep{RINEHART2021}. We propose to adapt the averaging technique to the test case. The aim of the averaging procedure is to smooth out local oscillations. These oscillations stem from the pore-scale geometry of the porous medium. However, by averaging, one also risks missing important quantitative changes in the behavior of the flow. 
If a flow, for instance, exhibits a fast decrease of the tangential velocity component on the interface, typically from high velocities in the free flow to very low velocities in the porous medium, averaging will smear out this contrast. However, if one chooses a large REV for volume averaging on the position of the interface for the Stokes flow, this effect is blurred.
On the other hand, choosing a small REV might introduce oscillations. 
To reduce these effects, we chose the exact size of the REV accordingly to the concrete test case. We will introduce the test cases in Section~\ref{chap:numericalexperiments} and give the REV used for each test case in Remark~\ref{rem:averaging}.

The third region (Fig.~\ref{fig:referencesol}) is the part of the free-flow region with a  distance of $\epsilon/2$ from the interface. Here, we average the pore-scale solution using $V(\vec a)= [a_1-\epsilon/2,a_1+\epsilon/2]\times[-\epsilon/4,a_2+\epsilon/4]$ as a REV. The averaging in this zone is necessary because we observe oscillations in the free flow close to the interface.
In the remaining part of the  free-flow domain (region 4) the solutions of the pore- and macroscale problems are compared directly.

\subsubsection{Quantification of the pore-macro mismatch}
The difference of processed pore-scale solution and macroscale solution is quantified in the porous-medium and free-flow domains separately. We evaluate the difference for each component of the velocity and the pressure. For such a scalar pore-scale quantity $\mathrm g^r$ and its corresponding macroscale quantity ($\mathrm g^\FF$ in the free flow and $\mathrm{g}^\PM$) belonging to the Stokes--Darcy problem with friction tensor $\vec\beta$ we define the relative error
functions
\begin{align}\label{eq:defDfg}
   \mathcal{D}^\FF_g= \frac{|\mathrm g^\FF-\mathrm g^r|}{\|\mathrm g^r\|_{L _2(\Omega_\FF^t)}}
\end{align}
in the free flow (in $\Omega_\FF^t$), and
\begin{align}\label{eq:defDpg}
    \mathcal{D}^\PM_g=\frac{|\mathrm g^\PM-\mathrm g^r|}{\|\mathrm g^r\|_{L _2(\Omega_\PM^{t})}}
\end{align}
in the porous medium (in $\Omega_\PM^t$). 
The variable $g$ can be $p$ for the pressure and $v_1$ or $v_2$ for the first or second velocity component, respectively. Thus $\mathcal{D}^\FF$ and $\mathcal{D}^\PM$ depend on the friction tensor $\vec \beta$. 
We aim to minimize the cost function that is obtained by adding the squares of the relative $L^2-$errors 
\begin{equation}
\begin{split}
 \begin{alignedat}{1} \label{eq:costfunction}
\mathcal{C}(\vec\beta)= &\|\mathcal{D}^\FF_{v_1}\|^2_{L_2(\Omega_\FF^t)}+\|\mathcal{D}^\FF_{v_2}\|^2_{L_2(\Omega_\FF^t)}+\|\mathcal{D}^\FF_{p}\|^2_{L_2(\Omega_\FF^t)}\\
+&\|\mathcal{D}^\PM_{v_1}\|^2_{L_2(\Omega_\PM^t)}+\|\mathcal{D}^\PM_{v_2}\|^2_{L_2(\Omega_\PM^t)}+\|\mathcal{D}^\PM_{p}\|^2_{L_2(\Omega_\PM^t)}\,,
\end{alignedat}
\end{split}
\end{equation}
to give greater weight to large deviations.
For a given set of boundary conditions, the evaluation of the cost function allows us to compare the suitability of different friction tensors as choices to obtain well-calibrated macroscale models. 

The cost function might be adapted to the specific test case, since the terms in \eqref{eq:costfunction} may differ greatly in magnitude. When a velocity component is very small, its relative error can become disproportionately large and dominate the cost function. In such situations, it may be necessary to omit that particular contribution from the cost function, as we will do in Section~\ref{subsubsec:comarison1Dflow}.

\subsection{Optimization procedure}\label{subsection:optimisation}
The optimization of the cost function will take place on different search spaces, that will be introduced in the next section. The second section considers the dual-annealing and the brute force algorithm.
\subsubsection{Search space}\label{subsubsec:searchspace}

We seek to minimize the cost function over search spaces constructed to include, at various levels, information obtained directly from the derivation of the stress–jump condition \citep{Angot_etal_17}.

In the derivation of the stress--jump conditions the friction tensor appears as an averaged version of a inverted permeability tensor in the transition region. 
As a consequence it is symmetric positive semidefinite. This property is also of importance for the well-posedness of the coupled problem. In contrast to the derivation of the stress--jump conditions we do not consider a transition region with continuously varying permeability. Nevertheless, it might be reasonable to assume that if we would consider a transition region, the structure of the inverted permeability tensor (principal axes, ratio of eigenvalues) would be similar in the transition region compared to the permeability in the porous medium. Since we consider a homogeneous porous medium with identical periodically distributed inclusions, we could deduce even more properties of the friction tensor. For example, in the case of an isotropic porous medium, the friction tensor is assumed to be scalar as the permeability.
We use the information obtained from the derivation on the supposed structure of the friction tensor to construct the search spaces. By relaxing the constraints on the structure of the admissible friction tensors, we test whether the assumptions on the structure of the friction tensor help us to find a well calibrated model or if by contrary the calibrated models exhibit a friction tensor with a different structure.

We now start by introducing different search spaces.
Let $\mathbf{K}$ be the constant permeability tensor defined in \eqref{def:permeability} and $\sigma_1,\sigma_2$ the corresponding eigenvalues. We denote a diagonal matrix with entries $d_1$ and $d_2$ by $\operatorname{diag}(d_1,d_2)$.
The first search space is 
\begin{equation}
\begin{split}
 \begin{alignedat}{1}\label{eq:defV1}
 V_1=\Bigg\{&\vec\beta\in\mathbb{R}^{2\times2}|\,\exists d\in \mathbb{R},\quad d>0:\\
 &\vec\beta=\dsp\frac{d}{\|\operatorname{diag}(\sigma_1,\sigma_2)\|_2}\mathbf{K}^{-1} \Bigg\} \,.
\end{alignedat}
\end{split}
\end{equation}\\
For this search space we will consider the coordinate space with respect to the basis vector.

In order to define the second search space, we note that permeability $\mathbf{K}$ is always orthogonally diagonalizable and so is $\mathbf{K}^{-1}$. We choose an orthogonal transformation $\mathbf{P}$, s.t. we get $\mathbf{K}^{-1}=\mathbf{P}\mathbf{D}\mathbf{P}^{-1}$, with $\mathbf{D}$ being a diagonal matrix. 
The second search space is defined as
\begin{equation}\label{eq:defV2}
\begin{split}
 \begin{alignedat}{1}
    V_2=\{\vec\beta\in\mathbb{R}^{2\times2}|\,\exists d_1,d_2\in \mathbb{R}, \quad d_1,d_2>0:\\\vec\beta=\mathbf{P}\operatorname{diag}(d_1,d_2)\mathbf{P}^{-1} \} \,.
\end{alignedat}
\end{split}
\end{equation}
For the first two search spaces $V_1$, $V_2$, symmetry and positive definiteness are enforced naturally. For our third search space 
\begin{align}\label{eq:defV3}
    V_3=\{\vec\beta\in\mathbb{R}^{2\times2}|\,\vec\beta=\vec\beta^\top\} \,
\end{align}
we do not enforce the positive definiteness assumption. Instead, in this case we add the penalization term 
\begin{align}\label{eq:penalisationterm}
     \textit{P}(\vec\beta)=\sum_{\sigma\in \rho(\vec\beta),\sigma<0}\frac{\sigma^2}{\sigma^2+0.1}
\end{align}to the cost function $\mathcal{C}$ that penalizes negative elements of the spectrum $\rho(\vec\beta)$ of the friction tensor. 
For a result that we obtained by the penalization approach we verify the positive semidefiniteness afterwards. 
In practice we will further restrict our search to a bounded subdomain of the corresponding search domain.

\subsubsection{Optimisation strategies}\label{subsubsect:optimisationstrategies}
We compare two search algorithms to improve the choice of metaparameters, like the number of function evaluations performed in one optimization process. If one algorithm leads to significantly better results than the other, this is an indication that one needs to change the metaparameters in the optimization algorithm that performs worse. This enhances the reliability of the calibration. We employ two optimization algorithms: the brute force search and the dual annealing algorithm. The brute force search corresponds to evaluating the cost function on a Cartesian grid and determining the minimum. The advantage of the brute force algorithm is its interpretability and its non-probabilistic nature. The dual annealing algorithm is a complex optimization scheme based on the simulated annealing algorithm \citep{XIANG1997}, whose goal is to mimic the cooling procedure of a material. The search space is explored in a probabilistic manner, allowing for leaving local minima. The dual annealing algorithm is a common choice for the calibration of complex systems \citep{SORITZ2022,HASSAN2022}. It is known for exploring the search space efficiently. By comparing the two algorithms, we can also improve parameter choices like the resolution in the brute force algorithm and the number of iterations in the dual annealing algorithm. We use the implementations of the dual annealing and brute force algorithm provided by the SciPy library \citep{VIRTANEN2020}.

To effectively apply the search algorithms we determine a subdomain included in the search space, excluding very large values of the friction tensor. More precisely we determine box constraints, i.e. intervals for each coordinate in our search space, to which we restrict the optimization. For the one-dimensional search space we start with the interval $[0,100000]$, perform a brute force search with 60 evaluations. We set the threshold as the optimal value of this search plus $0.02$. This choice is heuristic, based on the idea that we do not want to consider regions where the cost function is very large, but rather choosing the box constraints loosely. This is done by choosing a margin that is relatively large.  We determine the smallest interval containing all gridpoints for which the cost function is smaller than the threshold. However, the evaluation of the bounds should surpass the threshold. By iterating this approach, we can shrink the size of the domain, where the optimization will be conducted. We refer to this algorithm as the 'Box-constraint-algorithm'. Its pseudocode for the one-dimensional case is given in Appendix \ref{alg:boxsearch}. For the two-dimensional version of this algorithm, we perform the same procedure for a rectangle instead of an interval. The length of the Cartesian grid, where the cost function is evaluated, is chosen as $1.5$ times the upper value of the previous iteration in both components and $0.5$ times the value of the lower boundary. The grid has $81$ nodes. 

Extending this approach to the three-dimensional search space, would necessitate for the same resolution as in the two-dimensional case  $729$ evaluations of the cost function in every step of the iteration.
To avoid these high costs, we take the box-constraints for the friction tensor from the two dimensional case $[l_1,u_1]\times[l_2,u_2]$. The edges of $[l_1,u_1]\times[l_2,u_2]$   in $\mathbb{R}^2$ are $(l_1, l_2)$, $(l_1, u_2)$, $(u_1, l_2)$ and $(u_1, u_2)$. We recall that these are constraints for the eigenvalues of the friction tensor. We now calculate
\begin{align*}
  \mathbf{B1}&=\mathbf{P}\operatorname{diag}(l_1,l_2)\mathbf{P}^{-1}\,,\\
  \mathbf{B2}&=\mathbf{P}\operatorname{diag}(l_1,u_2)\mathbf{P}^{-1}\,,\\
  \mathbf{B3}&=\mathbf{P}\operatorname{diag}(u_1, l_2)\mathbf{P}^{-1}\,,\\
  \mathbf{B4}&=\mathbf{P}\operatorname{diag}(u_1,u_2)\mathbf{P}^{-1}\,.
\end{align*}
These values correspond to the edges of the transformed box-constraint. In order to determine the box-constraint for the three dimensional case, we now determine the maximum and minimum for each component of $\mathbf{B1}-\mathbf{B4}$ separately. The maxima are multiplied by $1.5$ and the minima by $0.5$ to reduce the risk of missing a minimum. These values are chosen as the box-constraints for the three-dimensional search space.  

For the test cases with isotropic porous medium, the non-diagonal elements in this approach would be equal to zero. Since we still want to explore the influence of a full symmetric tensor for isotropic porous media, we choose as bounds for the non-diagonal elements $\pm 0.5\operatorname{max}\{ u_1,u_2\}$.

The choice of the box constraints for the three-dimensional search spaces thus entail the two-dimensional boxes.

\begin{remark}\label{rem:CalibrationBJ}
    For the calibration of the Beavers--Joseph--Jones condition \eqref{eq:BJ}, we proceed analogously to the optimization in the one-dimensional search space replacing $V_1$ by $\mathbb{R}_+$. Here, we only use the dual annealing search strategy because in the calibration of the stress--jump conditions, it proved to work well.
\end{remark}

\subsection{Practical aspects of the implementaion}

\begin{figure}[htbp]
    \centering
    \includegraphics[width=0.9\linewidth]{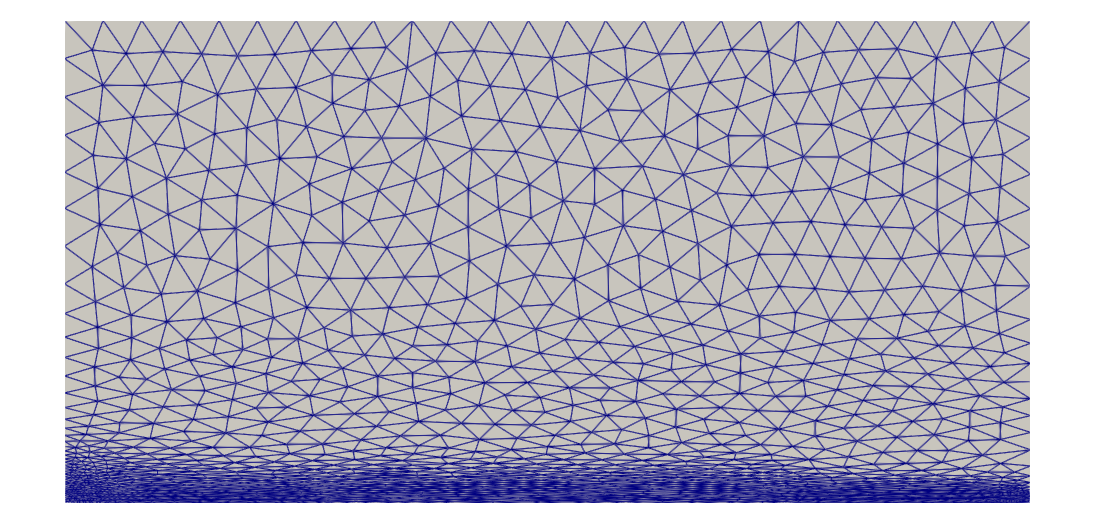}
    \caption{Mesh of $\Omega_\FF$, refined at the interface for Box-constraint-algorithm}
    \label{fig:refinedmesh}
\end{figure}

There are some practical aspects of the calibration that we will address next. 
Since we want to achieve a good exploration of the search space, we need to keep the computational costs of the evaluation of the cost function low. Because we are especially interested in the influence of the solution in the region close to the interface, we consider a macroscale mesh that is coarse away from the interface and gets more refined close to $\Sigma$ as can be seen in Fig.~\ref{fig:refinedmesh}. The pore-scale solution is interpolated on the macroscale mesh before the search starts. Since the reference solution partially consists of volume-averaged quantities, in the porous medium and free flow, the reference solution is not defined up to the boundary for except for region 4.

For the determination of the domain used for the optimization, a coarser grid, with maximal mesh size $0.1$ can be used (a typical length scale for the domains considered in our test cases is $1$). For the optimization schemes a mesh with maximal mesh size of below $0.02$ is used.

\subsection{Regional sensitivity analysis}\label{subsec:RSA}
For one isotropic porous medium we perform a regional sensitivity analysis (RSA). The aim of sensitivity analysis in general is to examine the influence of input parameters on the output of a model. In our case the input parameters are the entries of the friction tensor and the output is the value of the cost function. An RSA strategy aims at clarifying the influence of input parameters on the model that lead to output in a certain range (region)~\citep{PIANOSI2016}. We consider two methods proposed in \citet{WAGENER2001}, following the general guidelines for regional sensitivity \citep{PIANOSI2016}.  
The sampling has been done using a version of Sobol's sequence that is implemented in the Python library SAlib \citep{Iwanaga2022,Herman2017} for a subspace of $\mathbb{R}^2$. 
Both RSA techniques necessitate the following transformations of the obtained data.
The output values, i.e. the values of the cost function,  undergo two transformations: first, they are scaled so that larger numbers represent superior model performance, by subtracting for each sample the corresponding value of the cost function from the maximal value that the cost function takes for all samples. Then they are normalized by dividing by their total sum to ensure the sum of the all values of the transformed measure is equal to one. This leaves us with a ``transformed measure of performance'' \citep{WAGENER2001}. 

For the first RSA technique, in contrast to \citet{WAGENER2001}, we exclude all samples with values of the cost function above a threshold $\mathcal{s}$. For the remaining samples the input parameters are sorted with respect to their corresponding output into $10$ bins. For each of the bins the cumulative distribution of the parameters is plotted. A straight line indicates a less sensitive behavior with respect to the parameters, whereas different shaped curves are an indication for sensitivity. The restriction to samples with cost function below the threshold $\mathcal{s}$, was conducted to concentrate on those sets of parameters that are still in the proximity of the best performing set. This enables us, to choose for each bin a smaller range of parameters, so that the bins, corresponding with the best model performance don't contain samples, that are not at all acceptable for a calibration. The graphs for the cumulative distribution of the parameters in the different bins are Fig.~\ref{fig:cumdistfuncT1}, Fig.~\ref{fig:cumdistfuncT2}, and Fig.~\ref{fig:cumdistfuncT3}.

In the second RSA technique, the parameter space (the subspace of $V_2$ in which we perform the sampling) is divided into $30$ bins. The sum of the transformed measure in each bin is calculated, and the result is depicted in a histogram. In this way, the parameter density distribution ($D$) can be depicted. A uniform distribution indicates an insensitivity of the model with respect to the parameters in the chosen range. The graphs for the cumulative distribution of the parameters in the different bins are Fig.~\ref{fig:paramdistfuncT1}, Fig.~\ref{fig:paramdistfuncT2}, and Fig.~\ref{fig:paramdistfuncT3}.

\section{Numerical experiments}\label{chap:numericalexperiments}

In this section, we detail the calibration and validation of the stress--jump conditions \eqref{eq:SJic}. We start by describing different porous-medium geometries which we consider in Section~\ref{subsec:pore-geometries}. In  Sections~\ref{subsec:Testcase1}, \ref{subsec:Testcase2}, \ref{subsec:Testcase3} and \ref{subsec:Testcase4}, we present the flow scenarios for which calibration has been performed. The flow regimes were chosen as academical examples aiming at producing significant flow at the fluid--porous interface and in the porous medium. While parallel flow has been studied extensively, (e.g. \citep{EGGENWEILER2020,CARRARO2013,HERNANDEZ-RODRIGUEZ2022,valdes-paradaNovelOnedomainApproach2021})  cases with substantial normal flow component are less well studied in the literature \citep{YANG2019,SAHRAOUI1991}. 
In Section~\ref{subsec:results}, we discuss our results for the calibration. We compare the results for different optimization schemes, analyze the contributions to the cost function and the influence of the scale separation parameter $\epsilon$. We continue by comparing the stress--jump conditions with the classical interface conditions and performing a regional sensitivity analysis. Finally we turn to the case of flow tangential to the interface and discuss a formula that estimates the value of the friction tensor.

\begin{figure*}[htbp]
    \centering
    \includegraphics[width=0.25\textwidth]{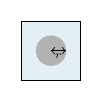}
    \hfill
    \includegraphics[width=0.25\textwidth]{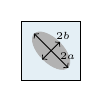}
    \hfill
    \includegraphics[width=0.25\textwidth]{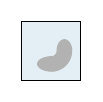}
    \caption{Solid inclusions: circular (left), ellipsoidal (middle) and rounded lune (right)}
    \label{fig:three_images}
\end{figure*}

\subsection{Pore geometries}\label{subsec:pore-geometries}

On the pore-scale the porous media examined here consists of a periodic repetition of a quadratic cell. This cell strictly contains inclusions that will form the solid matrix. For simplicity it suffices thus to describe the geometry in a unit cell, i.e. an open quadratic cell $Y=(0,1)^2$. The solid part of the unit cell $Y_s\subsetneq \overline{Y}$ is a closed subset. 
The domain of the porous medium resolved on the pore-scale is then defined by $\Omega^\epsilon_\PM=\Omega_\PM\setminus\cup_{r\in \mathbb{Z}^2}(Y_s\epsilon+r\epsilon+t)$ as depicted in Fig.~\ref{fig:geometries}, with $\epsilon$ being the scale separation parameter. The real number $t<\epsilon$ is chosen to ensure that the interface lies on top of the first row of inclusions.

 Due to the periodic structure, the permeability of the porous medium can be calculated by formula \eqref{def:permeability} provided from homogenization theory \citep{Hornung1997}. 
 We consider three different porous media. The permeability and porosity values are depicted in Table~\ref{tab:geometryparameter}.
The first type of inclusion is a circular shape of radius $r=0.25$. The corresponding porous medium is isotropic. 
The second shape is a tilted ellipse, with width $2a=0.8$ and height $2b=0.4$, leading to an orthotropic porous medium. Finally, a less symmetric inclusion is used that can be described as a rounded lune. Its parametrization is given in the Appendix \ref{secC}.
The interface is placed on top of the first row of inclusions.

\begin{remark}\label{rem:notionofoblique}
The notion of flow direction on the interface is ambiguous. On the pore-scale, the directions of the streamlines entering the porous medium can be used to define the flow direction. Alternatively, one can consider the direction of the averaged velocity. 
However, the choice of the averaging technique on the interface is not clear and could depend on the flow regime, as explained in Subsection \ref{subsec:ManufacturingOfReferenceSolutions}.
On the macroscale for two-domain models, the velocity fields are defined in the free flow and in the porous medium separately. The transition between these vector fields across the interface is not necessarily continuous, but depends on the interface conditions. Thus, also for the macroscale problem, the notion of flow direction at the interface is not obvious.

\end{remark}

 \begin{table}
\centering
\begin{tabular}{r| c c c}
\toprule
Parameter & circular   & ellipsoidal  &  lune-shaped\\
\midrule
Porosity  & 0.804& 0.615& 0.826\\
Permeability \\
$k_{11}$&$4.98\times 10^{-5}$ &$3.07\times 10^{-5}$&$4.74\times 10^{-5}$ \\
$k_{22}$&$4.98\times 10^{-5}$&$3.07\times 10^{-5}$  & $3.60\times 10^{-5}$ \\
$k_{12}$& 0&$-6.71\times 10^{-6}$ &$6.93\times 10^{-6}$\\
\bottomrule
\end{tabular}
\caption{ Parameters of porous media\label{tab:geometryparameter}}
\end{table} 

 \subsection{Test case 1}\label{subsec:Testcase1}
\begin{figure*}[htbp]
    \centering
    \includegraphics[scale=0.8]{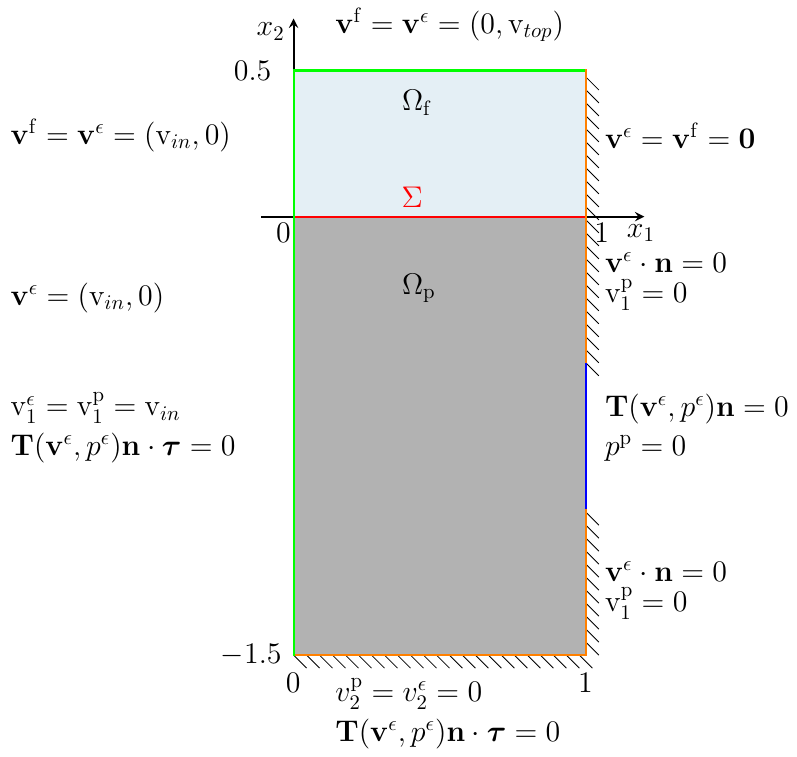}
    \caption{Boundary conditions on the pore scale and on the macroscale for Test case 1}
    \label{fig:T1}
\end{figure*}

In the first test case, we prescribe boundary conditions that lead to a velocity field exhibiting very different directions of flow, both in the porous medium and in the free flow. A case with different directions of the velocity field in the free-flow domain challenged efforts to calibrate the Beavers--Joseph condition \citep{STROHBECK2023}. In our case, the variance of flow directions is largest in the porous medium.
The flow on the interface is mainly directed towards the porous medium.
The boundary conditions of this first test case are shown in Fig.~\ref{fig:T1}.
 The tangential component of the inflow velocity at the left boundary is given by 
\begin{align}
\mathrm v_{in}(y)=\begin{cases}
    -4(y-0.5) & \text{if } y \geq 0.25, \\
    1 & \text{if } -0.25\leq y \leq 0.25,\\
    4(y+0.5)& \text{if }y\leq -0.25\,,
\end{cases}
\end{align}
with $y$ being the second coordinate according to Fig.~\ref{fig:T1}.
At the top boundary, we impose
\begin{align}
    \mathrm v_{top}=4(x-0.5)(x-0.5)-1
\end{align}
as the normal component of the velocity. We observe that the boundary conditions we set on the lower boundary of the porous medium introduce a large modeling error because the conditions on the pore and on the macroscale are equivalent only asymptotically. This leads to a cost function which values are dominated by the lower part of the porous medium. Since we are interested in the impact of the interface conditions, we exclude the lowest part of the porous medium from the calculation of the cost function. Namely, we set 
\begin{align*}
    \Omega_\PM^t\coloneqq\{&\vec x|\vec x\in\Omega_\PM, \\
    &\operatorname{dist}(\vec x,\partial (\Omega_\PM\cup\Omega_\FF\cup\Sigma)\geq\frac{\epsilon}{2} ;y>-1.2\}\,.
 \end{align*}
for Test case 1.

 \subsection{Test case 2}\label{subsec:Testcase2}
\begin{figure*}[htbp]
    \centering
    \includegraphics[scale=0.8]{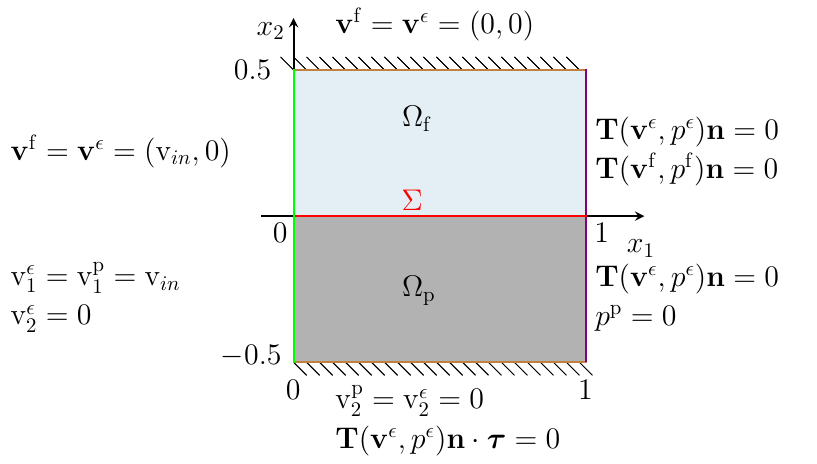}
    \caption{Boundary conditions on the pore scale and on the macroscale for Test case 2}
    \label{fig:T2}
\end{figure*}

 In the second test case, the flow is directed from the left boundary to the right boundary. The normal flow on the interface is directed towards the free flow.
 The boundary conditions are depicted in Fig.~\ref{fig:T2}.
 For the inflow velocity at the left boundary we impose
 \begin{align}
\mathrm v_{in}=\begin{cases}
    -4(y-0.5) & \text{if } y \geq 0.25\,, \\
    1 & \text{if } -0.25\leq y \leq 0.25\,,\\
    4(y+0.5)& \text{if }y\leq -0.25\,.
\end{cases}
\end{align}
Because the simulation domain is shorter compared to the first test case, we do not get outflow at the left boundary.

 \subsection{Test case 3}\label{subsec:Testcase3}
\begin{figure*}[htbp]
    \centering
    \includegraphics[scale=0.8]{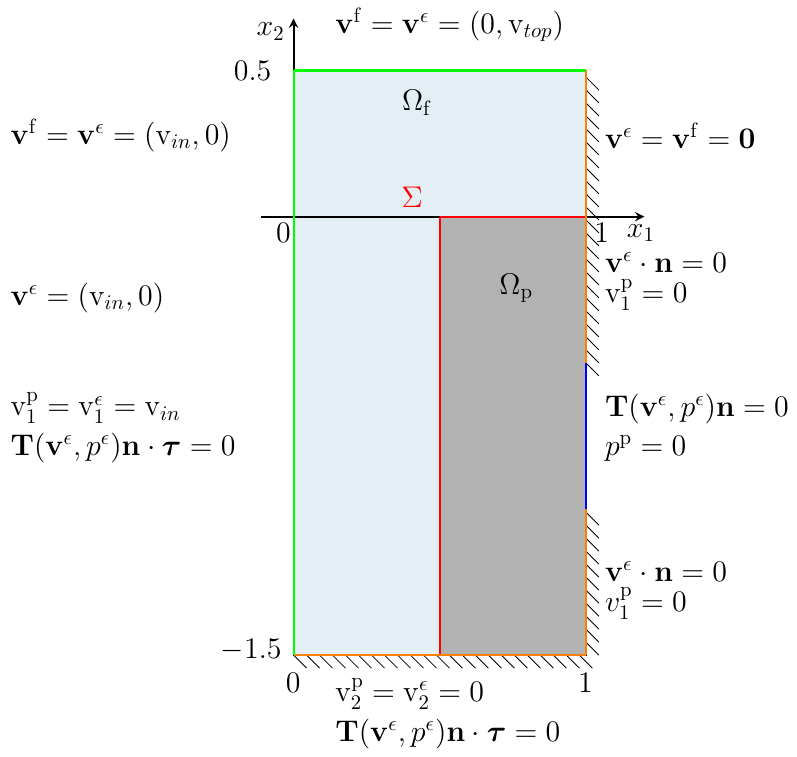}
    \caption{Boundary conditions on the pore scale and on the macroscale for Test case 3}
    \label{fig:T3}
\end{figure*}

In the third test case, we consider a rectangular porous medium that shares two interfaces with the free flow (Fig.~\ref{fig:T3}). In this test case, the interface is longer compared to the first two test cases. This allows for more complexity in the behavior of the flow on the interface. Furthermore, we can study effects that are due to changes in the geometry of the porous medium.

The horizontal velocity on the left boundary is set to
\begin{align}
v_{in}=\begin{cases}
    -4(y-0.5) & \text{if } y \geq 0.25, \\
    1 & \text{if } -1.25\leq y \leq 0.25,\\
    4(y+1.5)& \text{if }y\leq -1.25\,.
\end{cases}
\end{align}
At the top boundary,
\begin{align}
    v_{top}=4(x-0.5)(x-0.5)-1
\end{align}
is set as the normal component of the velocity.

  \subsection{Test case 4 - one-dimensional flow}\label{subsec:Testcase4}

\begin{figure*}[htbp]
    \centering
    \includegraphics[scale=0.8]{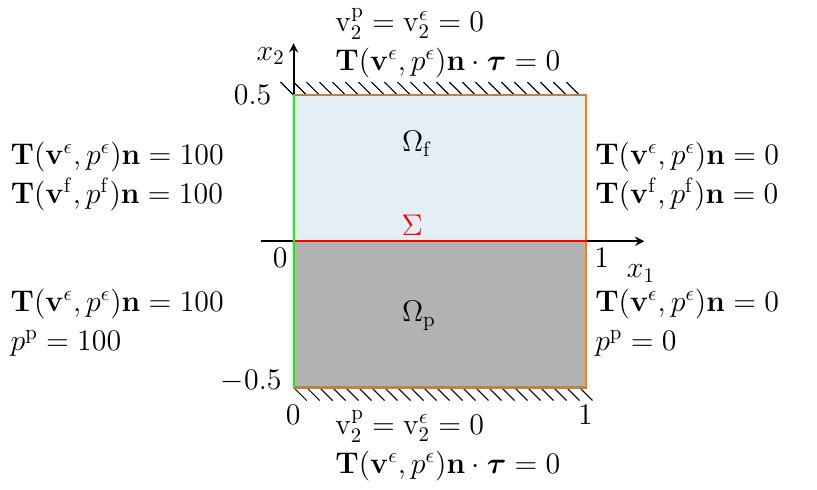}
    \caption{Boundary conditions on the pore scale and on the macroscale for Test case 4}
    \label{fig:T4}
\end{figure*}

  In the fourth test case, we compare our calibration strategy with an approximation formula for the case of isotropic porous media and scalar-valued friction tensor from \citep{Angot2021}. We set boundary conditions that extend the one-dimensional flow problem described by Angot in \citep[Sec.~4.2.1]{Angot2021} to a two-dimensional formulation. The flow is driven by a pressure difference between the left- and right boundary. The boundary conditions are given in Fig.~\ref{fig:T4}.

\begin{remark}\label{rem:averaging}
As explained in Section~\ref{subsec:ManufacturingOfReferenceSolutions}, we construct the reference solutions from the pore-scale solution by volume averaging in the regions 1, 2, and 3 depicted in Fig.~\ref{fig:referencesol}. In order to calculate a volume average in a point $\vec a=(a_1,a_2)$ we use as representative elementary volumes  $V1(\vec a)=\bigl[a_1-\epsilon/2,\; a_1+\epsilon/2\bigr]\times\bigl[-\epsilon/2,\; a_2+\epsilon/2\bigr]$ and $V2(\vec a)=\bigl[a_1-\epsilon/2,\; a_1+\epsilon/2\bigr]\times\bigl[-\epsilon/4,\; a_2+\epsilon/4\bigr]$. In region 1 we use $V1$ in region 3 we use $V2$. For region 2 the choice of REV depends on the test case. For Test case 1 and 3 we use $V1$ and  for Test case 2 and 3 $V2$. This choice is based on the experience, that the respective choice in the calibration leads to a smaller value of the cost function, i.e. the reference solution aligns better with the corresponding macroscale solution.

\end{remark}

\subsection{Results}\label{subsec:results}
Here, we present the findings obtained in the calibration process. 
We start by providing the obtained optimal values of the cost function concerning different search spaces and search algorithms in Section~\ref{subsubsec:Comparisonoptallpm}. We analyze the optimal results for different methods. 
In Section~\ref{subsubsec:flowregimeanalysis} we consider the difference between the reference solution and the macroscale simulation, interpreting the spatial behavior of this function and qualitatively explaining its behavior. 
Section~\ref{subsubsec:modelerror} then estimates the modeling error arising solely from the use of Darcy’s equation by comparing pore‑ and macroscale problems in the porous medium, with suitable interface boundary conditions replacing the coupling conditions \eqref{eq:SJic}. 
In Section~\ref{subsubsec:Comparisonclassicalic} we compare for one flow regime and all three types of inclusions the calibrated Stokes--Darcy system with stress--jump conditions \eqref{eq:SJic} to the Stokes--Darcy problem with classical coupling conditions \eqref{eq:classicalic}, where the dimensionless number $\alpha_\BJ$ is calibrated analogously to the friction tensor (see Remark \ref{rem:CalibrationBJ}).
Section~\ref{subsubsec:RSA} presents a regional sensitivity analysis (RSA) for calibrating the friction tensor for circular inclusions, shown in Fig.~\ref{fig:three_images} (left).
Finally, for the circular inclusions, Section~\ref{subsubsec:comarison1Dflow} investigates the applicability of the calibration results obtained by \citet{Angot_etal_21} for one-dimensional flow.

\subsubsection{Comparison of optimization strategies and different porous media}\label{subsubsec:Comparisonoptallpm}
\begin{figure}[htbp]
    \centering
    \includegraphics[width=0.92\linewidth]{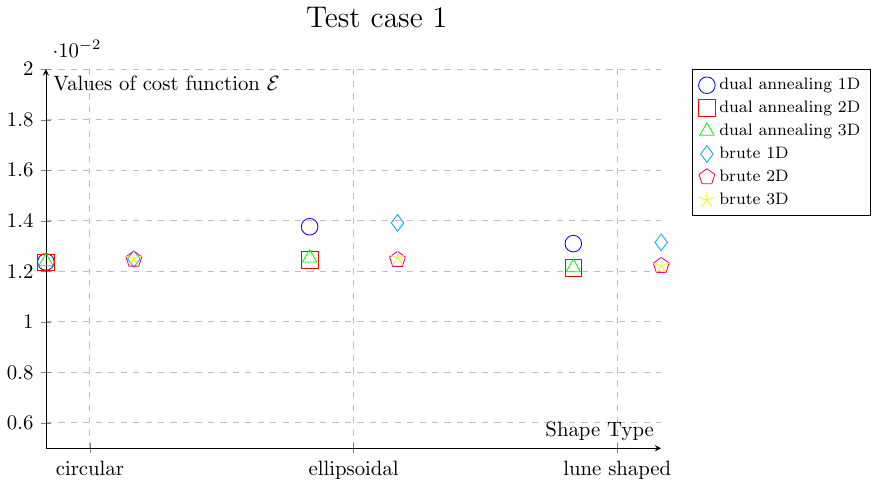}
    \includegraphics[width=0.97\linewidth]{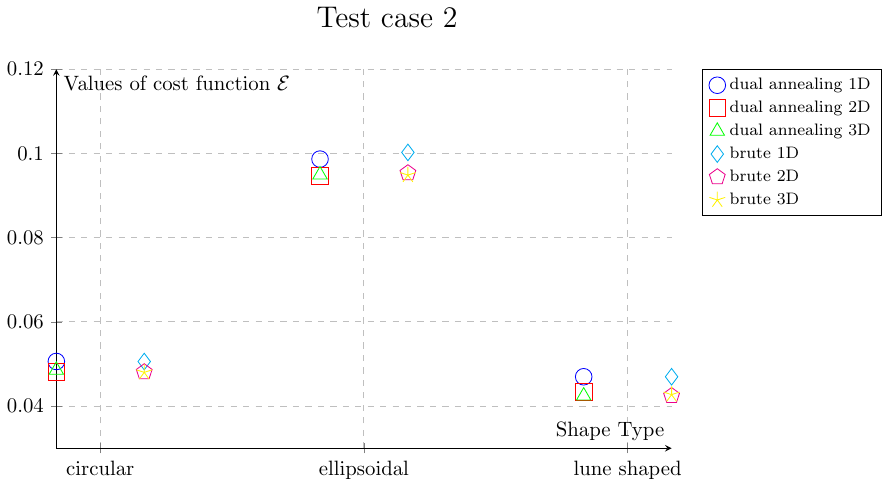}
    \includegraphics[width=0.92\linewidth]{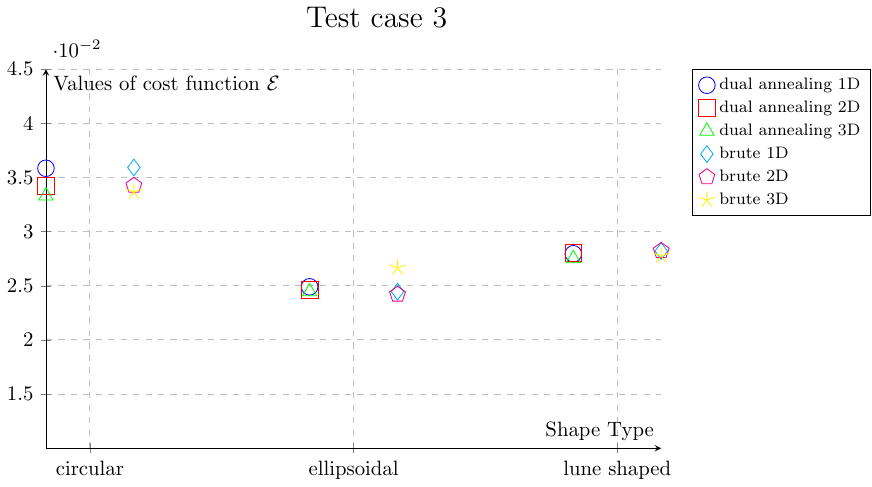}
    \caption{Comparison of different optimization strategies}
    \label{fig:compsearchalg}
\end{figure}

We chose the parameters of the optimization algorithms in such a way that they necessitate a similar amount of evaluations of the cost function. This allows for a fair comparison. In Table~\ref{tab:results} we give the best values for the friction tensor found in all optimization schemes.
In Fig.~\ref{fig:compsearchalg}, we compare the dual annealing algorithm and the brute force algorithm for the different search spaces introduced in Section~\ref{subsubsec:searchspace}. 

\begin{table*}[htbp]
\centering
\caption{Test case results: the best friction tensor values obtained by any of the optimization schemes: $[\beta_{11},\beta_{22},\beta_{12}]$ }\label{tab:results}
\begin{tabular}{lccc}
\toprule
Test case & circular & ellipsoidal & lune-shaped \\
\midrule
Test case 1 & [2.01256, 2.90539, 0.00000] & [2.69854, 2.69854, 0.75082] & [3.82514, 5.42785, -0.97912] \\
Test case 2 & [1.81848, 2.70539, -0.58756] & [3.15100, 3.15100, -0.26944] & [1.88744, 3.74706, -0.99300] \\
Test case 3 & [2.25361, 7.94698, 0.50143] & [0.26652, 0.26652, 0.07368] & [2.63519, 11.16680, -1.02800] \\
\bottomrule
\end{tabular}
\end{table*}

\begin{figure}[htbp]
    \centering
    \includegraphics[width=0.95\linewidth]{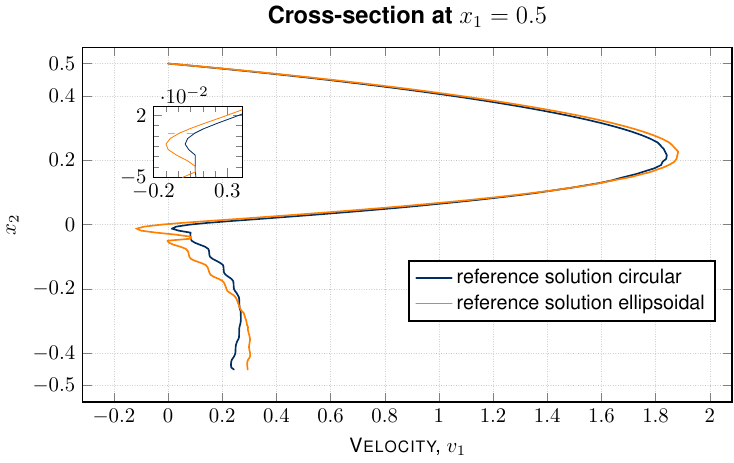}
    \caption{Comparison of the first velocity component of the reference solution at $x_1=0.5$ for circular and ellipsoidal inclusions in the second test case}
    \label{fig:BJ-crossection}
\end{figure}

In the first test case, we see that the optimal error values are of similar magnitude for all porous-media types.
For the second test case, one can observe that for the ellipsoidal inclusions the  error is significantly larger compared to the other two types of inclusions. This difference is probably due to the shape of the pores at the interface. In contrast to the other two inclusion types, this shape limits the directions of flow at the interface. Since this effect is only true on the pore scale, this induces stronger oscillations as can be seen in Fig.~\ref{fig:BJ-crossection}, by comparing the reference solutions for circular and ellipsoidal inclusions.
In the case of lune shaped inclusions, the dual-annealing algorithm performs slightly better than the brute force algorithm for two- and three-dimensional search spaces.

In the third test case, for the circular test case, the use of a one-dimensional search space leads to slightly worse results compared to larger parameter spaces. For the ellipsoidal and lune shaped inclusions, the results are almost the same.

To summarize our findings, we see, that the differences between the optimal values of the cost function for different search spaces are not large. In many cases, even the one-dimensional search space is sufficient.
This observation is a strong indication that the calibration of the stress--jump conditions can be reduced to a one-dimensional problem in practice.

Other choices for the tensor used as a basis vector in the construction of the one-dimensional search space (instead of the inverse permeability), in our experience can lead to significantly worse results though. For example in Test case 3 the choice of the permeability tensor leads to an minimal error of approximately $0.04$ for dual-annealing and brute force search algorithm. This is evidence supporting the assumptions on which the derivation of the stress--jump conditions is based, namely the modeling of the boundary layer by a transition zone.

\begin{figure*}[htbp]
    \centering
        \includegraphics[width=0.95\linewidth]{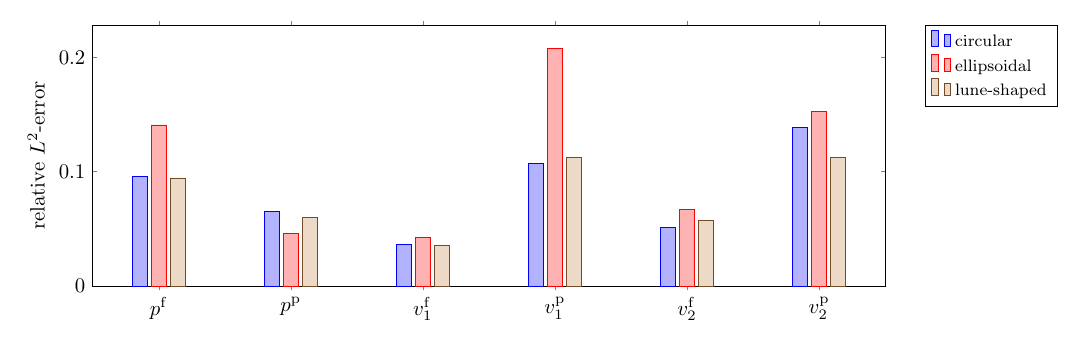}
    \caption{Comparison of the $L^2$-errors for different quantities for the  second   test case for the Stoke--Darcy problem with stress--jump conditions}
    \label{fig:quantitycompT2}
\end{figure*}

We also give the $L^2$-error values obtained for the calibrated systems (i.e. the $L^2$-norm of the difference between the  Stokes--Darcy model with the stress-jump conditions  and the reference solution for the friction tensor that corresponds to the smallest value of the cost function). For the first test case we see in Fig.~\ref{fig:quantitycompT2} that for most quantities the order of magnitude of the error values for different inclusion types is comparable. Only in the free flow, the error in the velocity components is larger for the lune-shaped inclusions. 

\begin{figure*}[htbp]
    \centering
        \includegraphics[width=0.95\linewidth]{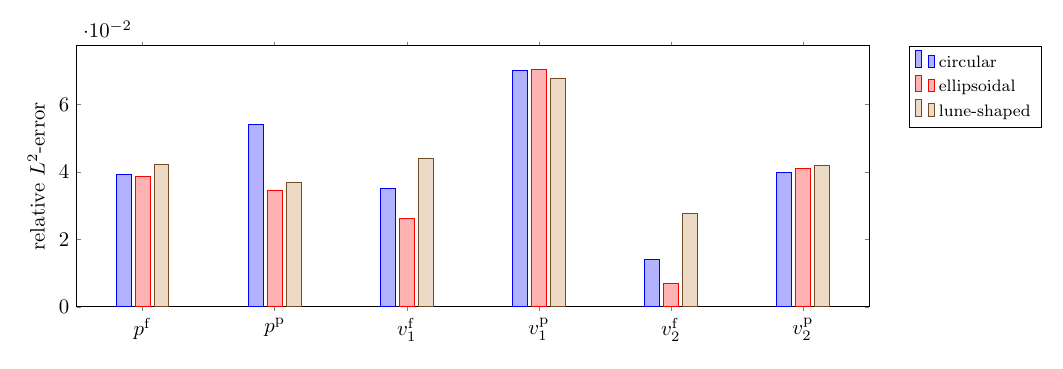}
    \caption{Comparison of the $L^2$-errors for different quantities for the first  test case for the Stokes--Darcy problem with stress--jump conditions}
    \label{fig:quantitycompT1}
\end{figure*}
For the second test case we see similar results in Fig.~\ref{fig:quantitycompT2}. However, for ellipsoidal inclusions the error values for the pressure in the free flow and for the first velocity component in the porous medium are substantially larger. As explained before, by examining the first velocity component in the porous medium, we see that close to the interface, the reference solution still exhibits an oscillation, responsible for this augmented value.

\begin{figure*}[htbp]
    \centering
    \includegraphics[width=0.95\linewidth]{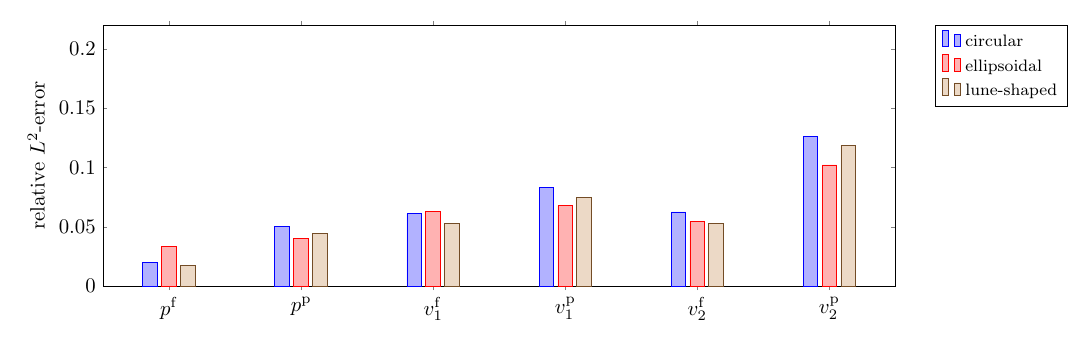}
    \caption{Comparison of the $L^2$-errors for different quantities for the  third  test case with stress--jump conditions}
    \label{fig:quantitycompT3}
\end{figure*}

For the third test case the $L^2$-errors are depicted in Fig.~\ref{fig:quantitycompT3}. Here, the error values for each inclusion type are of the same order of magnitude. 
By comparing the results from Fig.~\ref{fig:quantitycompT1}, Fig.~\ref{fig:quantitycompT2}  and Fig.~\ref{fig:quantitycompT3} we observe that the distribution of the error varies significantly across the quantities between the test cases. 

\subsubsection{Analysis of the contributions to the cost function}\label{subsubsec:flowregimeanalysis}
In this section, we consider only isotropic porous media based on circular inclusions. This case will be used to study the spatial distribution of the difference between macroscale and reference solutions, thus allowing us to understand better how different choices in the pore- and macroscale models (boundary conditions, type of inclusion) and in the construction of the reference solution (averaging technique applied for pore-scale solutions) contribute to the values of the cost function.

In the following, we choose the friction tensor that leads to the smallest value of the cost function from both dual-annealing and brute force algorithm, for one- two- and  three-dimensional search spaces \eqref{eq:defV1}-\eqref{eq:defV3} and work with the resulting calibrated model.

\begin{figure*}[htbp]
    \centering
    \begin{minipage}[t]{0.49\textwidth}  
        \centering
\includegraphics[width=\textwidth]{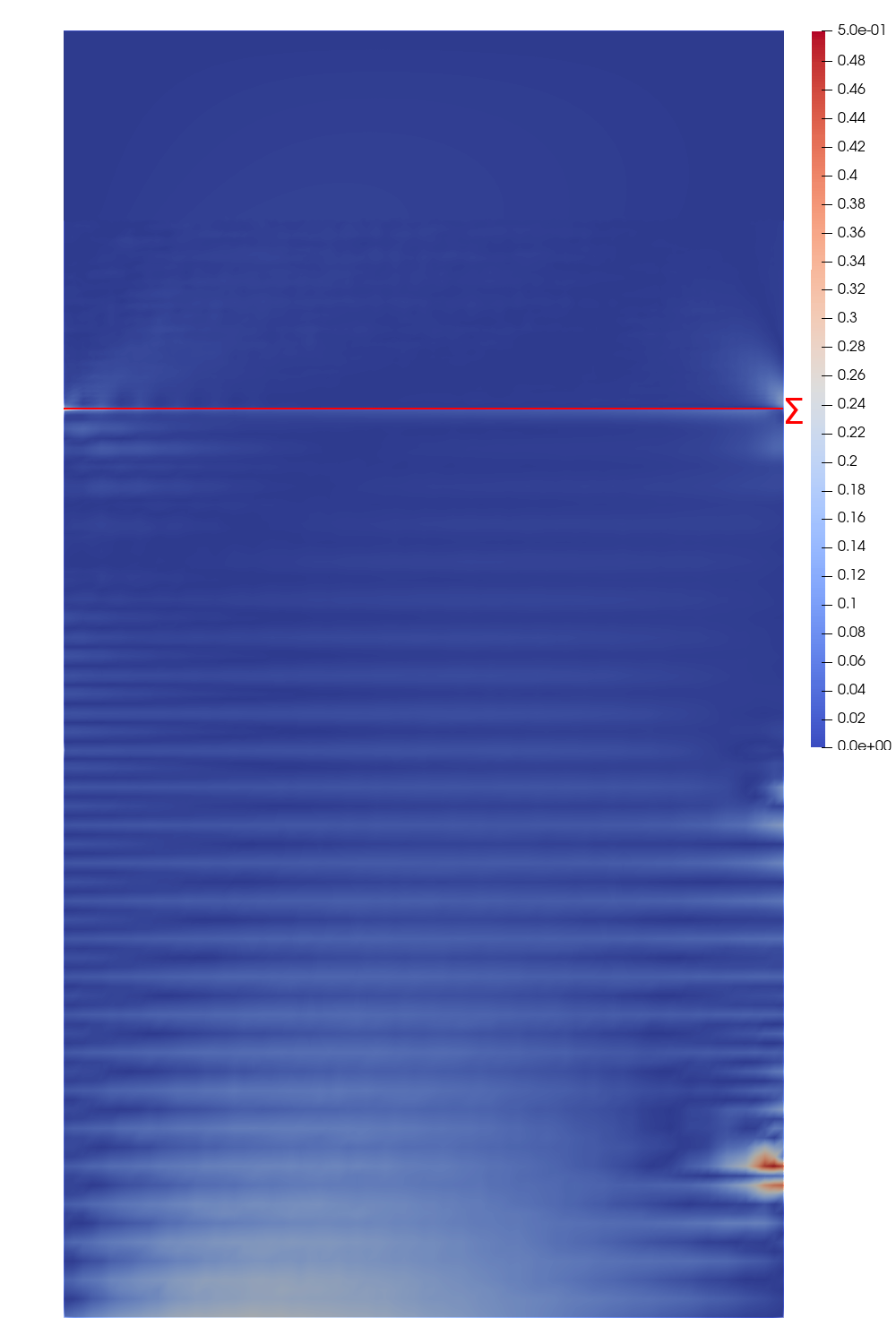}
    \end{minipage}\hfill
    \begin{minipage}[t]{0.49\textwidth}
        \centering
                \includegraphics[width=\textwidth]{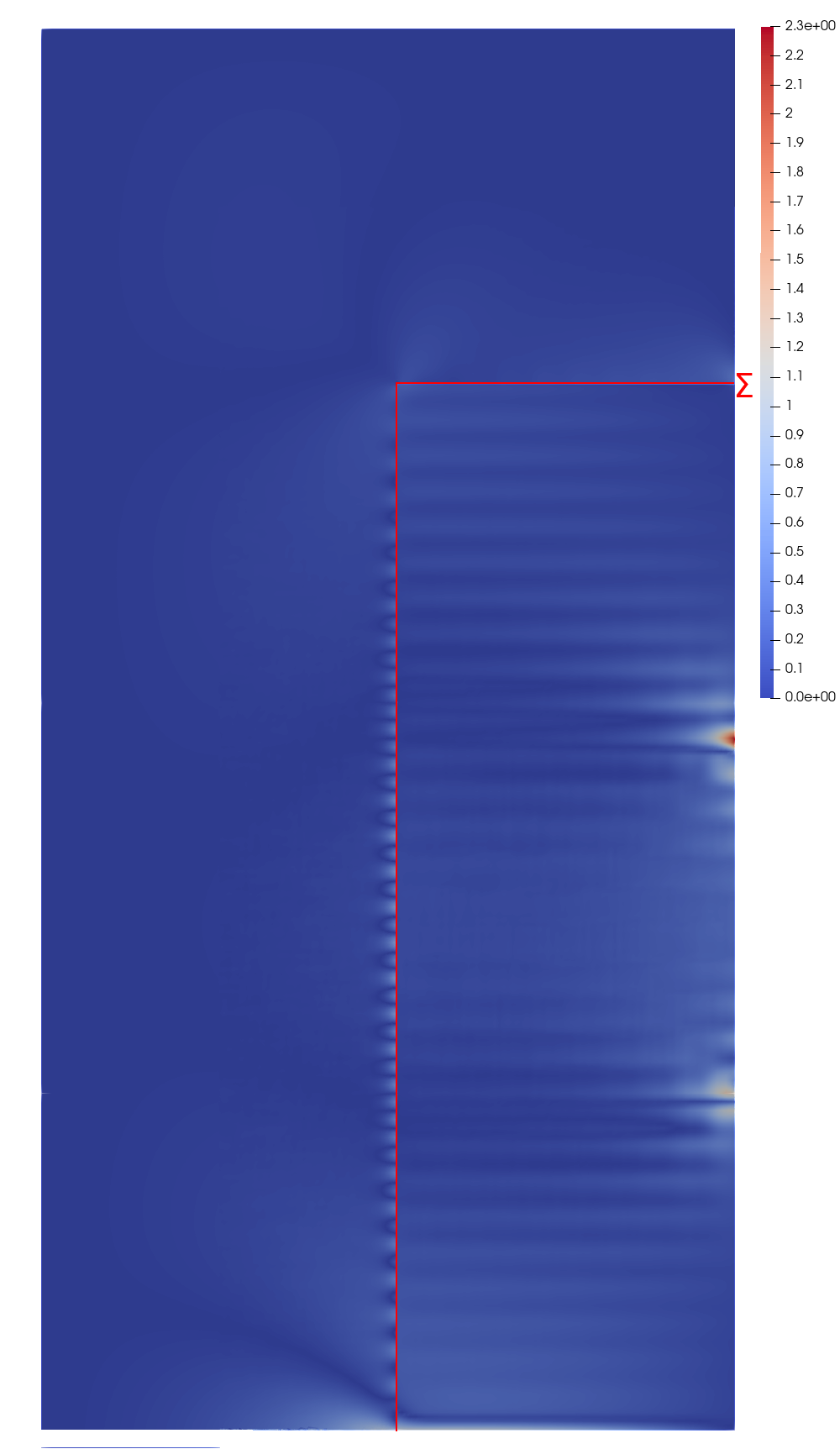}
    \end{minipage}
    \caption{Spatial distribution of $\mathcal{D}^\PM, \mathcal{D}^\FF$ with $\epsilon=0.05$, for for Test case 1 (left), Test case 3 (right)}
    \label{fig:spatialerrordist}
\end{figure*}

We plot the functions $\mathcal{D}^\PM_{v_1}$ and $\mathcal{D}^\FF_{v_1}$, introduced in \eqref{eq:defDpg}, \eqref{eq:defDfg}.
The result is depicted in
 Fig.~\ref{fig:spatialerrordist} for Test case 1 and 3. In Fig~\ref{fig:spatialerrordisteps}, we depict the results for Test case 2 with two different values of the scale separation parameter $\epsilon$.

These plots have a similar appearance for the other quantities (second velocity components and pressure). 
In all plots, we see that the largest differences are attained in the porous medium. This is also reflected in the comparison of the $L^2$-errors for the first velocity components in the free flow and the porous medium in Fig.~\ref{fig:quantitycompT1}, \ref{fig:quantitycompT2} and \ref{fig:quantitycompT3}. 
In the porous medium, there are periodical stripes that indicate oscillations. These stem  from the reference solution. By choosing a larger REV one could decrease this error. However, this would increase the computational cost of the manufacturing of the reference solution, decrease the part of domain, where a reference solution can be calculated and increase the risk of smearing out the flow profile for the reference solution.

For the first test case for the calibration, as explained in Section~\ref{subsec:Testcase1}, we do not consider the lowest part of the porous medium, hence the bottom of Fig.~\ref{fig:spatialerrordist} is at height $y=-1.2$. As one can see in the graphic there is a significant mismatch between reference solution and macroscale model at the bottom of the domain, that is already truncated. This mismatch is due to the choice of boundary conditions in the porous medium for the macroscale and pore-scale models. Boundary conditions for the Stokes equations in porous media and the corresponding Darcy equations can be equivalent only in an asymptotic manner. Because we are interested in the error introduced by the interface conditions we chose to truncate the domain for the calibration. At the edges of the outflow, we notice considerable errors that stem from channel flow between the right boundary of the domain and the last column of inclusions. For a smaller scale separation parameter $\epsilon$, this effect would be less pertinent. A similar but less pronounced artifact can be found at the most left and most right part of the interface, both in the free flow and in the porous medium. For this test case the error at the interface is concentrated in a very small stipe in the porous medium with width way below $\epsilon$. Thus, they reflect on a mismatch between the macroscale model and the reference solution in the boundary layer region that is to be expected, when comparing a coupled Stokes--Darcy problem with constant permeability with a pore-scale based reference solution. Such a macroscale model, in fact cannot resolve the behavior of the fluid in the boundary layer.

For the second test case, the spatial distribution of the error is depicted in Fig.~\ref{fig:spatialerrordisteps} for two different sizes of $\epsilon$. In the vicinity of the interface on the left, the same artifact can be observed as in the first test case. In the free-flow domain there is a broad stripe where $\mathcal{D}^\FF$ is elevated. Because of the size of this zone, that is higher than $4\epsilon$, this error rather seems to be a direct consequence of the coupling conditions than due to a local flow occurring on a scale that is too small to be represented in an up-scaled flow. We will detail the influence of the size of the scale separation parameter $\epsilon$ in Section~\ref{subsubsec:modelerror}

When considering the third test case (Fig.~\ref{fig:spatialerrordist}, right hand side), we notice that at the edges of the outflow as in the first test case, an increase in the error values can be observed. Furthermore, directly at the vertical part of the interface in the free-flow domain, $\mathcal{D}^\FF$ exhibits oscillations. As in the second test case in some parts of the free-flow domain adjacent to the interface the error is also larger.

\subsubsection{Influence of different scale separation parameters on calibrated model}\label{subsubsec:modelerror}

In the previous section, we saw that some of the errors can be traced back, to flow phenomena at pore-scale that should become less significant for smaller $\epsilon$ and should shrink in size (with the geometry). 
This is, because the theoretical derivation of Darcy's law is based on a small scale separation parameter for volume averaging \citep{WHITAKER1986a} or is conducted for the limit $\epsilon\to 0$  in homogenization \citep{Hornung1997}. Furthermore as the friction tensor $\vec \beta$ is non-dimensional, we expect, that once we calibrated our model, a change of $\epsilon$ would not necessitate a new calibration.
To provide evidence for these claims we study in the following the second test case with a smaller $\epsilon$. 
This section consists of two parts. In the first part, we consider only a porous medium domain of the same shape as the porous medium in Test case 2. For the scale separation parameter $\epsilon$ we choose the values 0.05 and 0.025. We consider a flow regime similar to the one in the porous medium in Test case 2. 
In the second part, we consider Test case 2 for the same two values of $\epsilon$. We use for the friction tensor the optimal result from all our calibration approaches applied for $\epsilon=0.05$.
The relative $L^2-$errors for the two values of $\epsilon$ obtained in the first part (i.e. only for porous media) give us a reference for the corresponding $L^2-$errors for the coupled problem. If the friction tensor does not depend on $\epsilon$ we should observe the same behavior for the values of the errors of the coupled problem for a change of $\epsilon$. 

\begin{figure*}[htbp]
    \centering
    \begin{minipage}[t]{0.49\textwidth}
        \centering
\includegraphics[width=\textwidth]{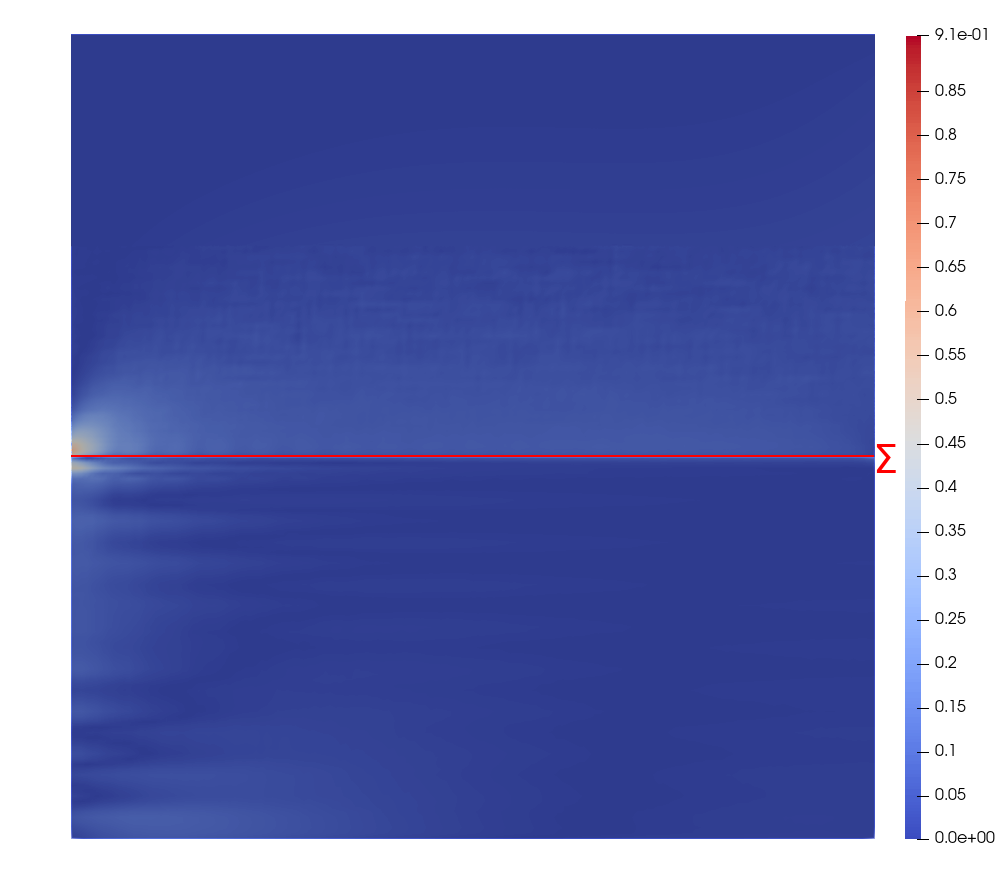}
    \end{minipage}\hfill
    \begin{minipage}[t]{0.49\textwidth}
        \centering
\includegraphics[width=\textwidth]{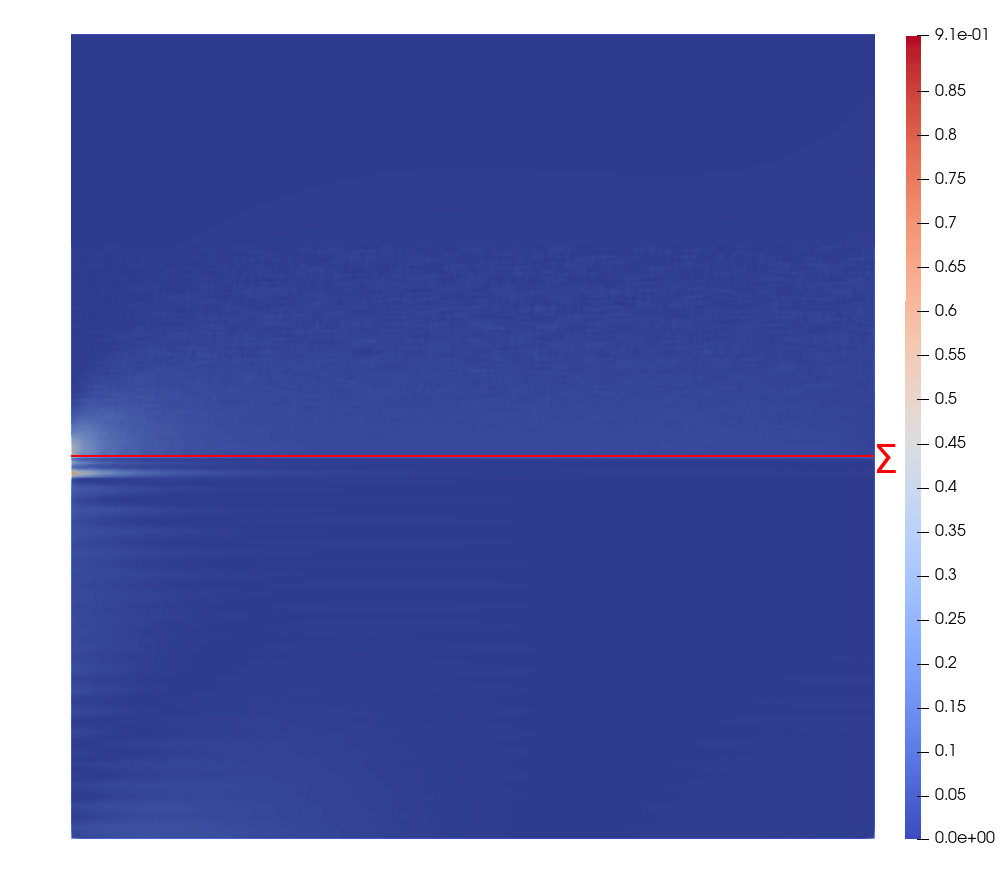}
    \end{minipage}
    \caption{Spatial distribution of $\mathcal{D}^\PM, \mathcal{D}^\FF$ for Test case 2;  $\epsilon=0.05$ (left) and $\epsilon=0.025$ (right)}
    \label{fig:spatialerrordisteps}
\end{figure*}

\begin{figure}[htbp]
    \centering
\includegraphics[width=0.8\linewidth]{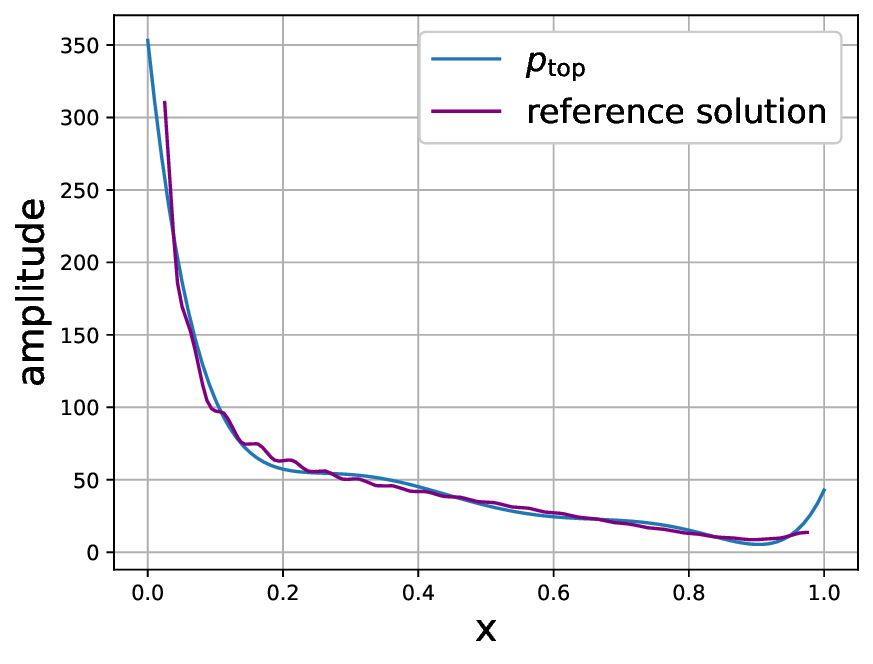}
    \caption{Reference solution and $p_\text{top}$ at $y=0.00625$}
    \label{fig:least-squarefit}
\end{figure}

Starting with the first part, we consider only the porous medium part of the simulation domain. To ensure a regular triangulation, our simulations take place in the domains $\Omega_{\PM}^t=[0,1]\times[-0.5,0.00625]$ and $\Omega_{\PM}^{t,\epsilon}=\Omega_{\PM}^t\setminus\cup_{r\in \mathbf{Z}^2}(Y_s\epsilon+r\epsilon+s)$. We use the same notation as in paragraph \ref{subsec:pore-geometries}. We impose at the top boundary of this new domain the boundary condition
\begin{align}
    \mathbf{T}(\vec v^\epsilon,p^\epsilon)\vec n=(p_{\text{top}},0)^\top
\end{align}
for the pore-scale model
and 
\begin{align}
    p^\PM=p_{\text{top}}
\end{align}
for the macroscale model.
The quantity
\begin{equation}
\begin{split}
 \begin{alignedat}{1}\label{eq:leastsquarepre}
       p_{\text{top}}=&353.19-4419.01x+25909.024x^2\\
    &-76009.79x^3+ 116582.64x^4\\
    &-89554.10x^5+27180.85x^6
\end{alignedat}
\end{split}
\end{equation}
is determined by a least-squares fit for a polynomial of degree six from the reference solution of the full problem at height $y=0.0625$. The comparison of the data to the fitted curve is depicted in Fig.~\ref{fig:least-squarefit}.
We calculate the reference solution as for the coupled flow problem. Now, in the domain $(-0.025,0.975)\times(-0.45,-0.01875)$, the cost function, i.e. the square of the relative $L^2$-errors can be computed. 
For the same boundary conditions we also consider the case of a smaller scale separation parameter $\epsilon=0.025$.

\begin{figure}[h]
    \centering
\includegraphics[width=0.95\linewidth]{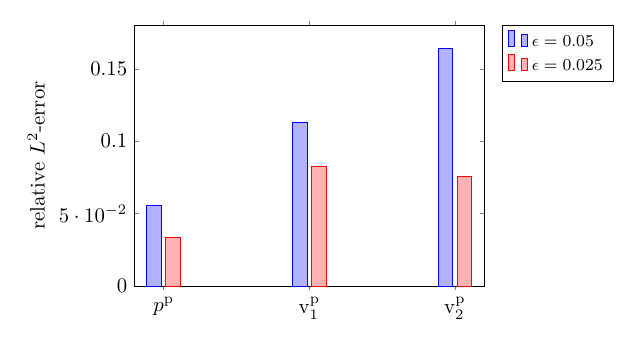}
    \caption{Relative $L^2$-errors for different sizes of $\epsilon$ for a flow in an isotropic porous medium}
    \label{fig:modelingerrorPM}
\end{figure}

The results are presented in Fig.~\ref{fig:modelingerrorPM}. First of all, for a smaller size of $\epsilon$, the difference between reference solution and the solution at macroscale decreases. 
This is in accordance with the theoretical derivation of Darcy's law. 
Secondly, by comparing the values of the $L^2$-error for $\epsilon=0.05$ of the considered flow problem in the porous medium with the corresponding quantities of Test case 2 for circular inclusions depicted in Fig.~\ref{fig:quantitycompT2}, we see that the order of magnitude for these errors is the same. This indicates that the calibrated Stokes--Darcy model with stress--jump conditions indeed leads to reasonable results.

\begin{figure*}[htbp]
    \centering
\includegraphics[width=0.95\linewidth]{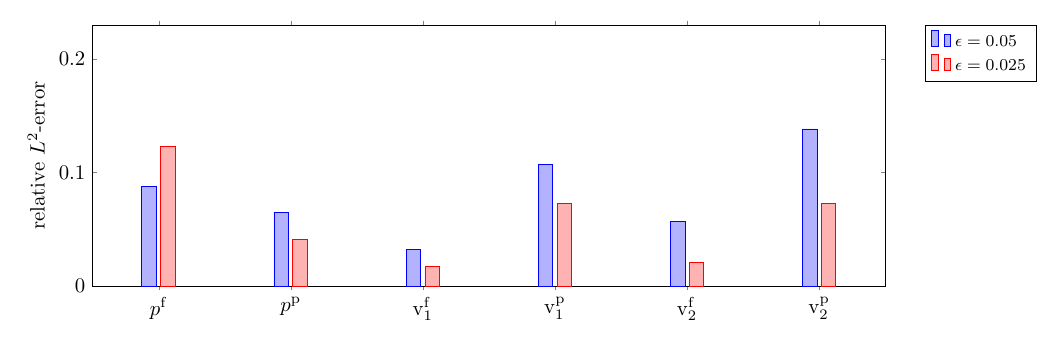}
    \caption{Relative $L^2$-errors for different sizes of $\epsilon$ for Test case 2 and circular inclusion with different scale separation parameter $\epsilon$}
    \label{fig:modelingerrorcoupled}
\end{figure*}

Now we turn to the second part of this section.
We examine how the calibrated coupled models behave when we reduce the size of $\epsilon$. As friction tensor we choose the friction tensor $\vec\beta$ obtained from our calibration procedure for $\epsilon=0.05$.
We want to verify, if this friction tensor is in fact non-dimensional, i.e. if for a different $\epsilon$ the changed $\sqrt{\mathbf{K}}$ in \eqref{eq:SJ} is sufficient to obtain a reasonably calibrated model for smaller pore-sizes. In practice this can be useful, to calibrate the model for a larger $\epsilon$, where the calculation of a reference solution is feasible and using the result to simulate flow for a porous medium with smaller pores.
The results depicted in Fig.~\ref{fig:modelingerrorcoupled} show a decrease in the error values for a smaller scale separation parameter. Only for the pressure in the free flow an increase of the error can be observed. This increase is due to a drastic rise of the pressure close to the interface in the porous medium in the reference solution, that is not observed for the coupled Stokes--Darcy models, because they only handle boundary layer effects by the coupling conditions and do not resolve the behavior inside the boundary layer. Since the pressure difference between porous medium and free flow for smaller $\epsilon$ is bigger, the induced error is bigger as well. 
The behavior of the errors related to the other quantities, especially those for pressure and velocity inside the porous medium show that the calibration for this test case is compatible with a change of the scale separation parameter.

We notice in Fig.~\ref{fig:spatialerrordisteps} that, indeed, the artifacts in the left part close to the interface decrease, as well as the periodic patterns inside the porous medium.

\subsubsection{Comparison with the classical interface conditions}\label{subsubsec:Comparisonclassicalic}
\begin{figure}[htbp]
    \centering
    \includegraphics[width=0.95\linewidth]{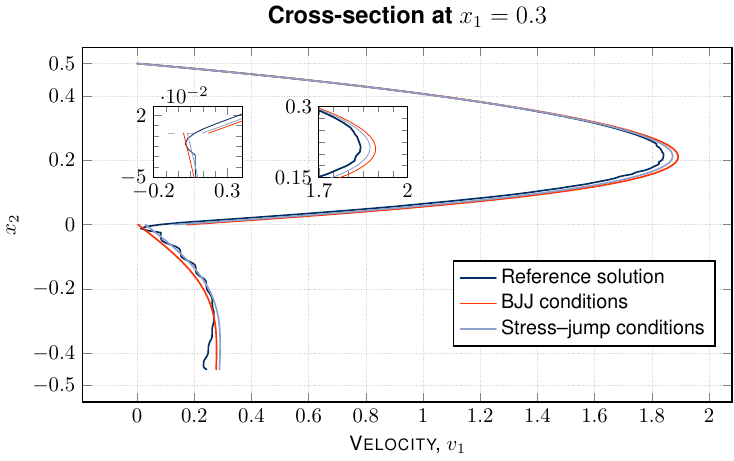}
    \caption{First velocity components at $x_1=0.3$ for the reference solution, and the calibrated Stokes--Darcy problem with the Beavers--Joseph--Jones and stress--jump conditions for circular inclusion, $\epsilon=0.05$}
    \label{fig:BJcompSJcirc}
\end{figure}

\begin{figure*}[htbp]
    \centering
    \includegraphics[width=0.95\linewidth]{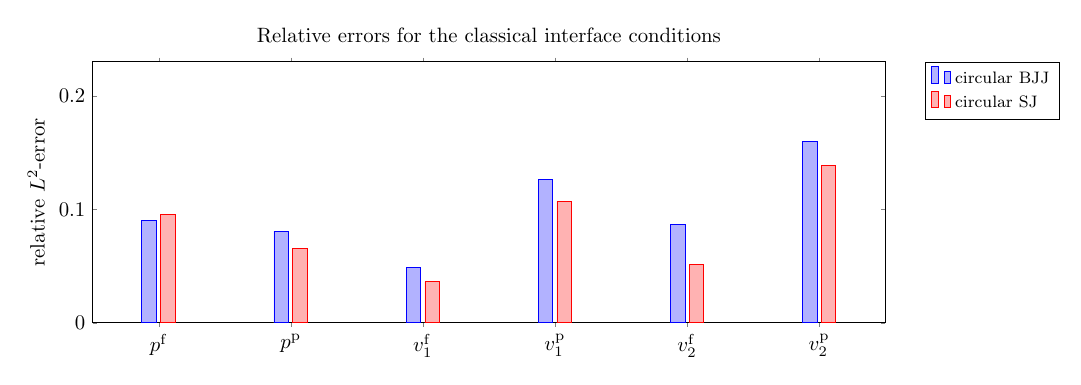}
    \includegraphics[width=0.95\linewidth]{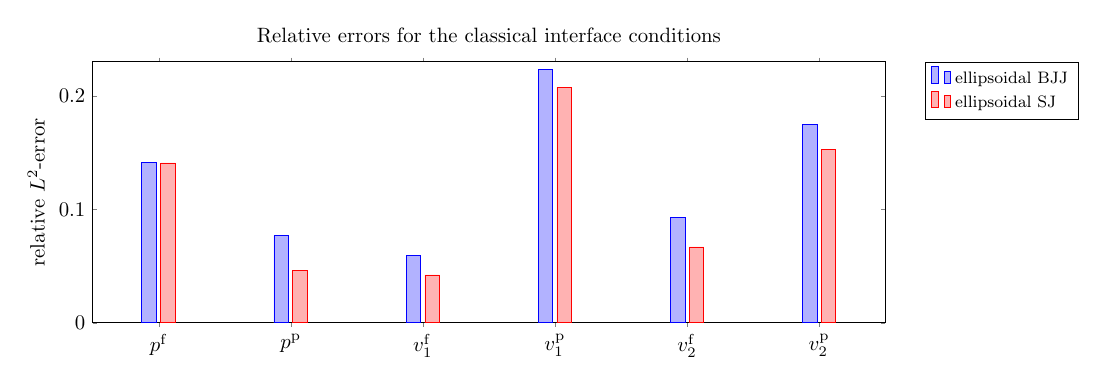}
    \includegraphics[width=0.95\linewidth]{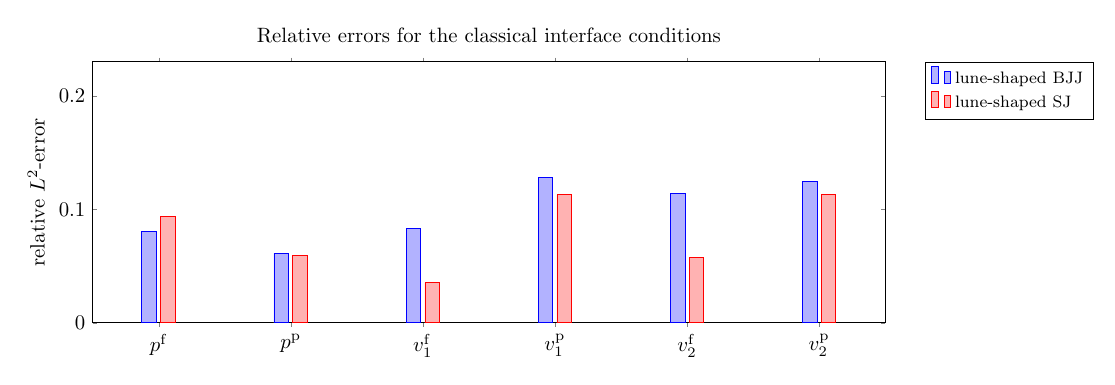}
    \caption{Comparison of relative $L^2$-errors for circular, ellipsoidal and lune-shaped inclusion for the calibrated Beavers--Joseph--Jones (BJJ)and stress jump Stokes--Darcy model  model (SJ)}
    \label{fig:BJquantitative}
\end{figure*}

For the second test case, we compare the results obtained for the stress--jump conditions \eqref{eq:SJic}, with those of the classical interface conditions \eqref{eq:classicalic}. The Beavers--Joseph parameter $\alpha_\BJ$ is determined by minimization of the cost function using the dual-annealing algorithm, which proved to work well in the calibration for the stress--jump model \ref{rem:CalibrationBJ}. In Fig.~\ref{fig:BJquantitative} we compare the relative $L^2$-errors obtained for the best performing calibrated stress--jump model with the corresponding errors for the calibrated model with Beavers--Joseph--Jones coupling conditions for each of the geometries. We note, that for most quantities, the calibrated stress-jump model outperforms the classical Beavers--Joseph--Jones model. This result is a strong indicator, that for coupled flow in cases where the flow in the porous medium is of the same order of magnitude as in the free flow, the stress--jump conditions are suitable and might even outperform the classical interface conditions.
To further visualize this we provide a plot of a cross-section at $x_1=0.3$ for Test case 2 with circular inclusions (Fig.~\ref{fig:BJcompSJcirc}).
Furthermore, we notice from Fig.~\ref{fig:BJquantitative} that the order of magnitude of the error values for the stress--jump model and the Beavers--Joseph--Jones model for most cases stays the same. This  can be explained by the fact, that many contributions to the overall error are independent of the choice of the interface conditions, as we explained in Section~\ref{subsubsec:flowregimeanalysis}. For example, the error values for the ellipsoidal test case are higher especially in the first velocity component, because the geometry of the pores at the interface induces strong oscillations as can be seen in \ref{fig:BJ-crossection}.


\subsubsection{Regional sensitivity analysis}\label{subsubsec:RSA}
\begin{figure}[htbp]
    \centering
    \includegraphics[width=0.9\linewidth]{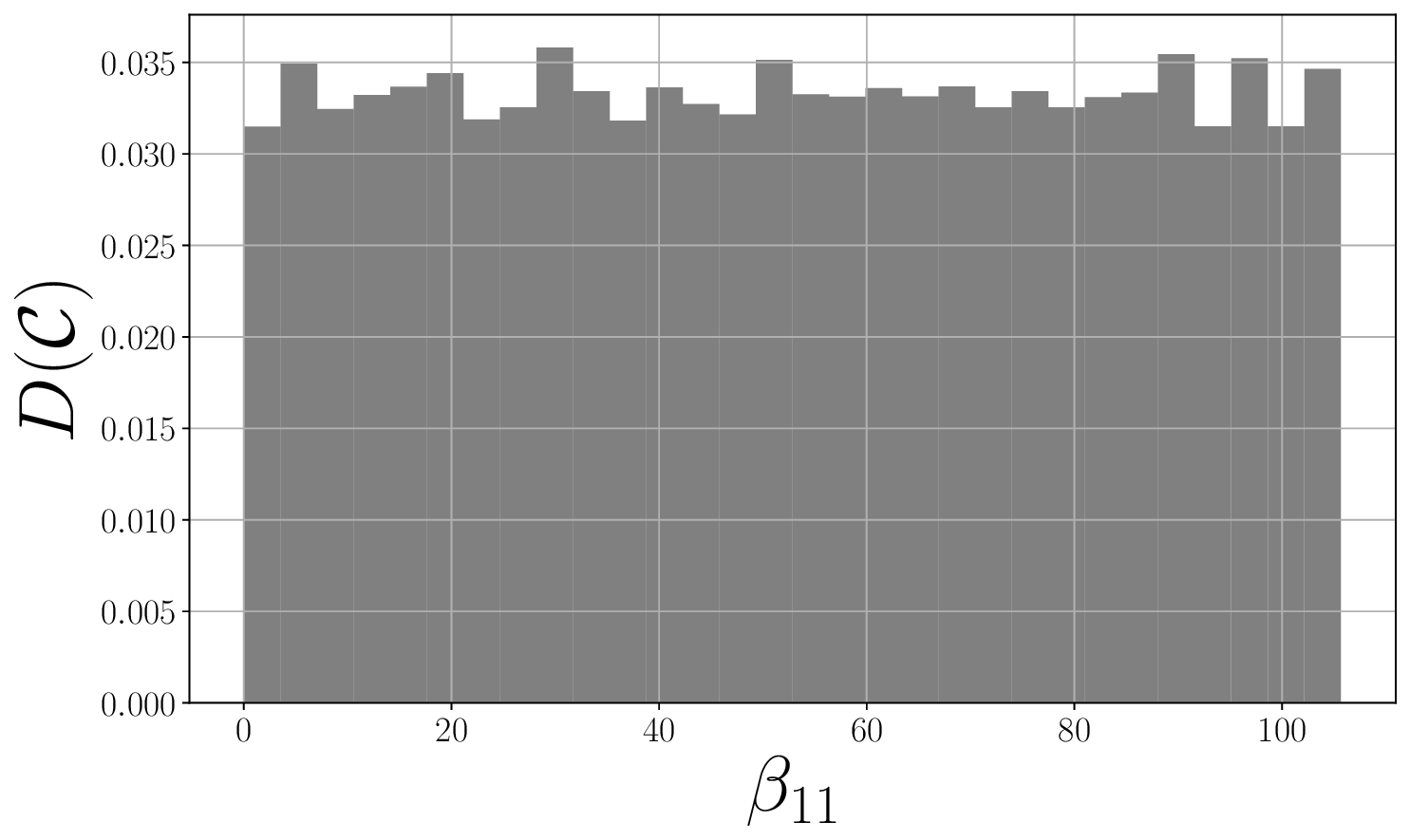}

\includegraphics[width=0.9\linewidth]{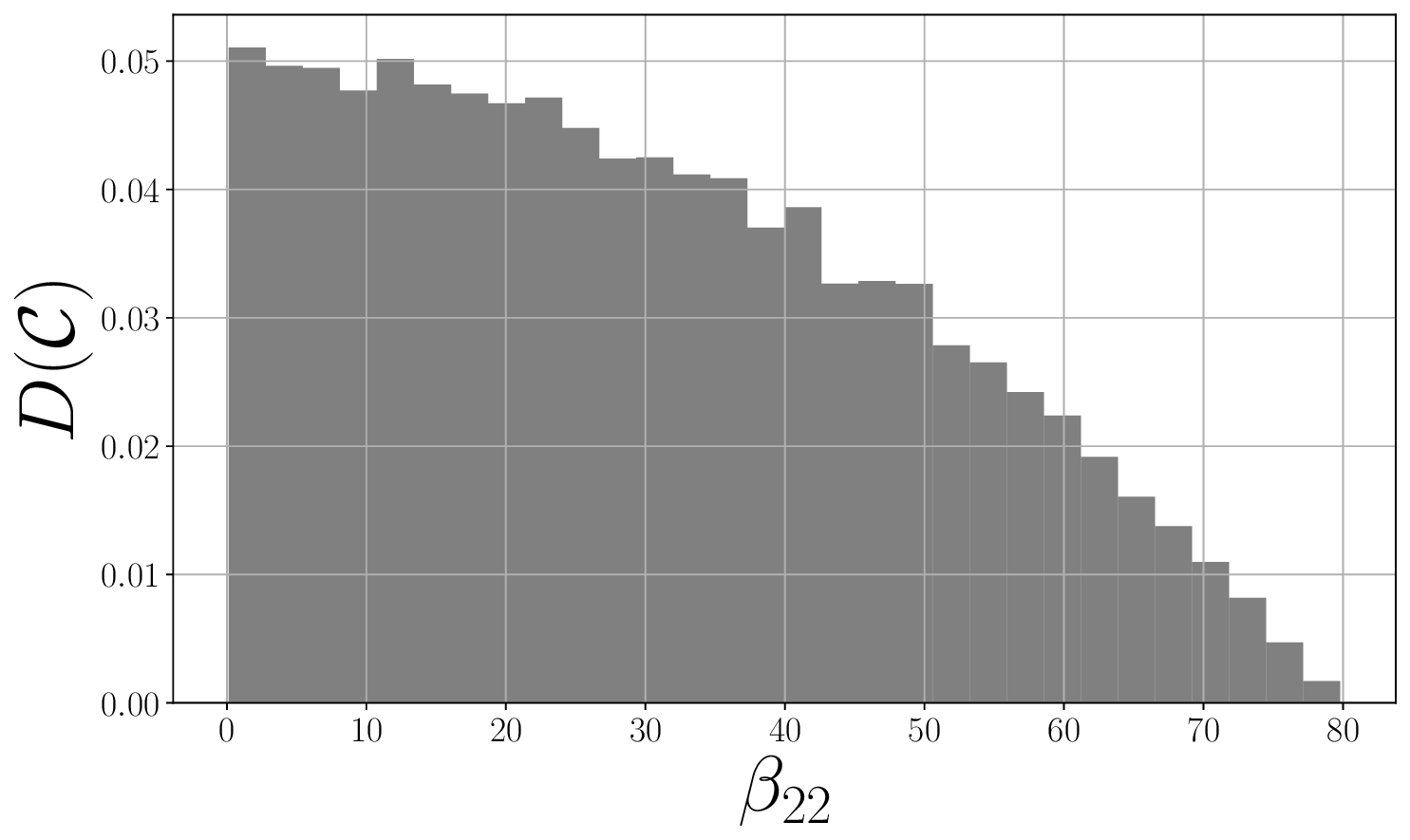}

    \caption{Parameter density distribution ($D$) for the two dimensional search space for Test case 1}
    \label{fig:paramdistfuncT1}
\end{figure}

\begin{figure}[htbp]
    \centering
    \includegraphics[width=0.9\linewidth]{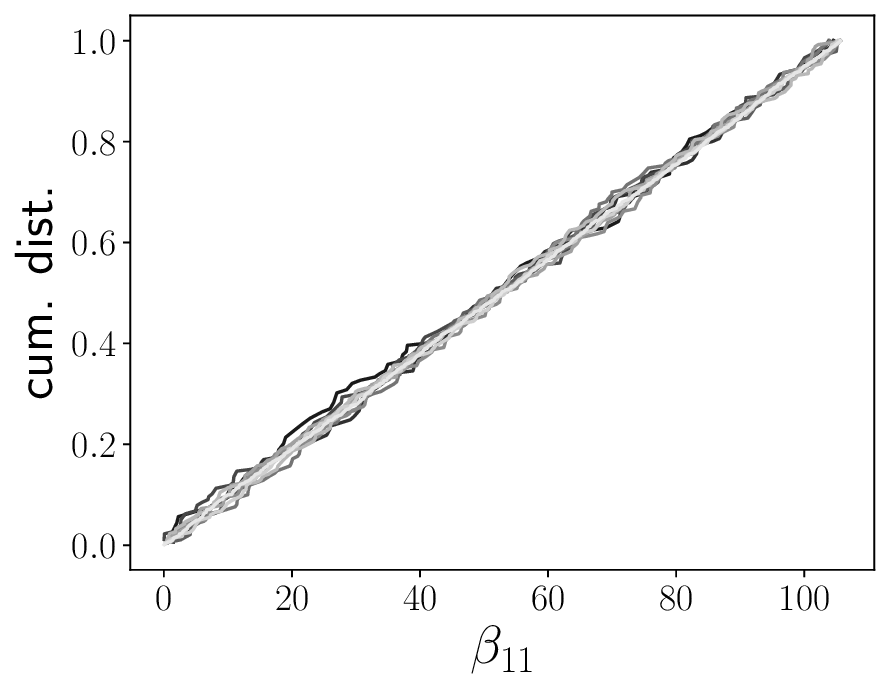}

    \includegraphics[width=0.9\linewidth]{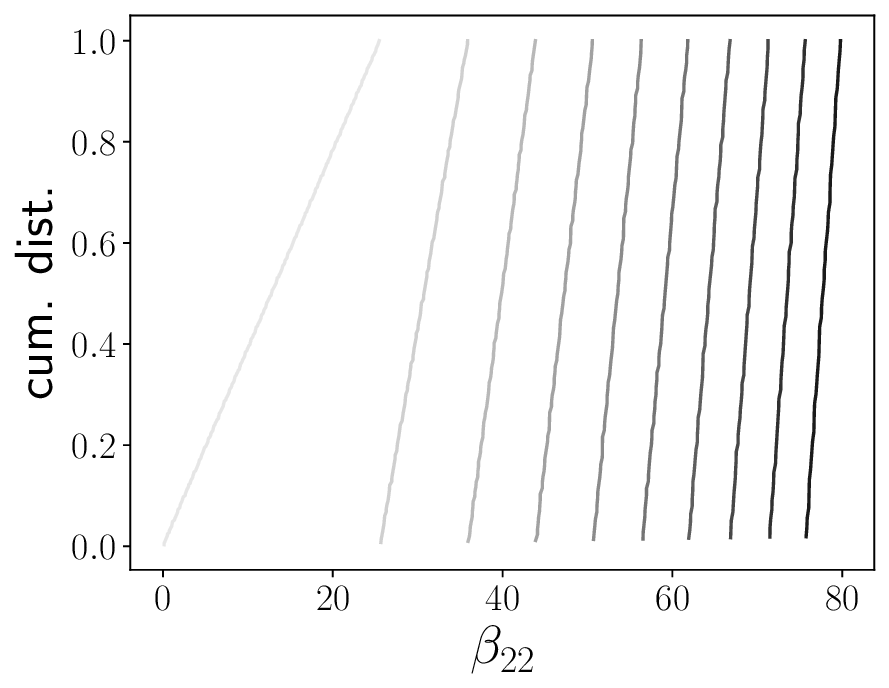}

    \caption{Cumulative distribution function  for the two dimensional search space for Test case 1}
    \label{fig:cumdistfuncT1}
\end{figure}

\begin{figure}[htbp]
    \centering
    \includegraphics[width=0.9\linewidth]{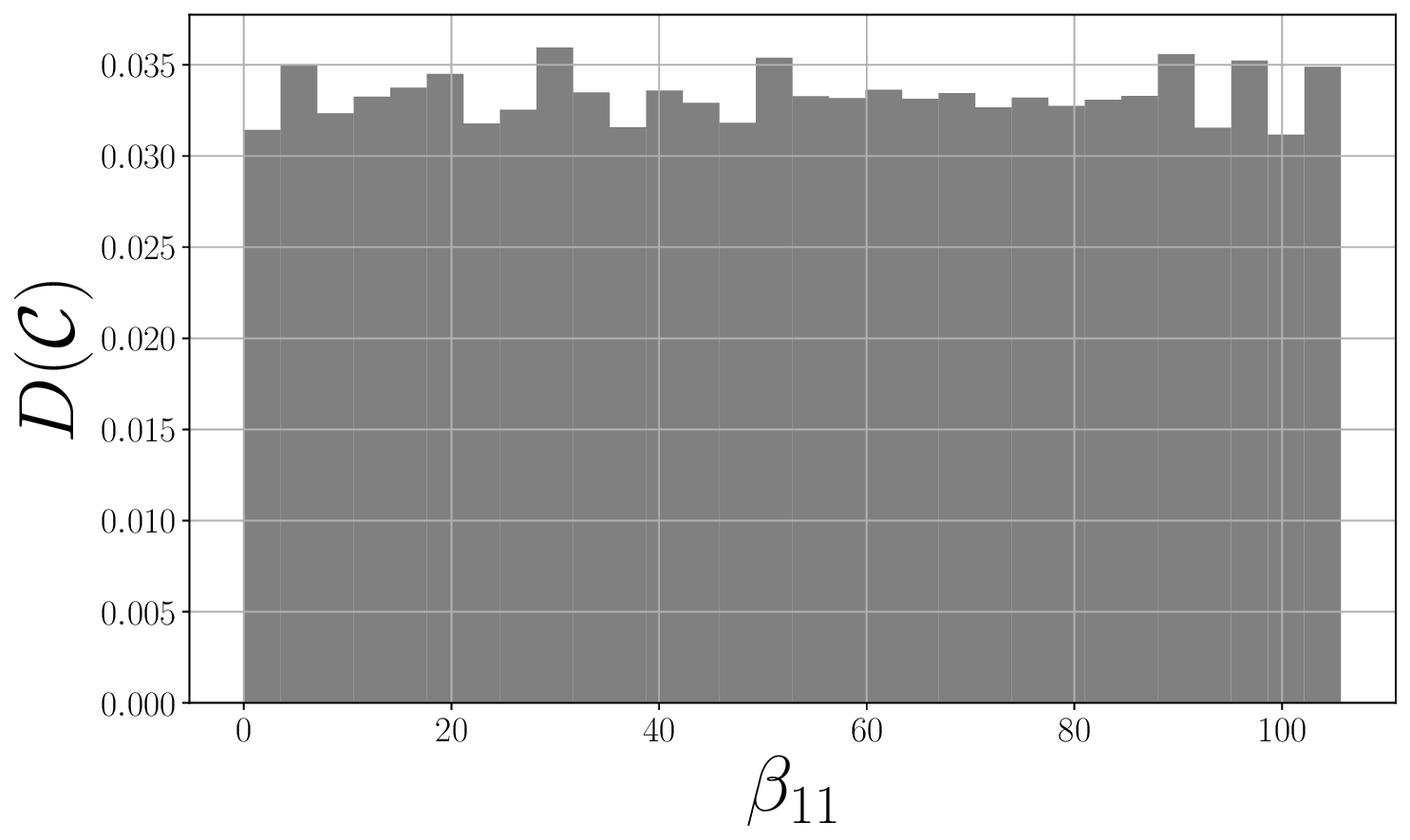}

\includegraphics[width=0.9\linewidth]{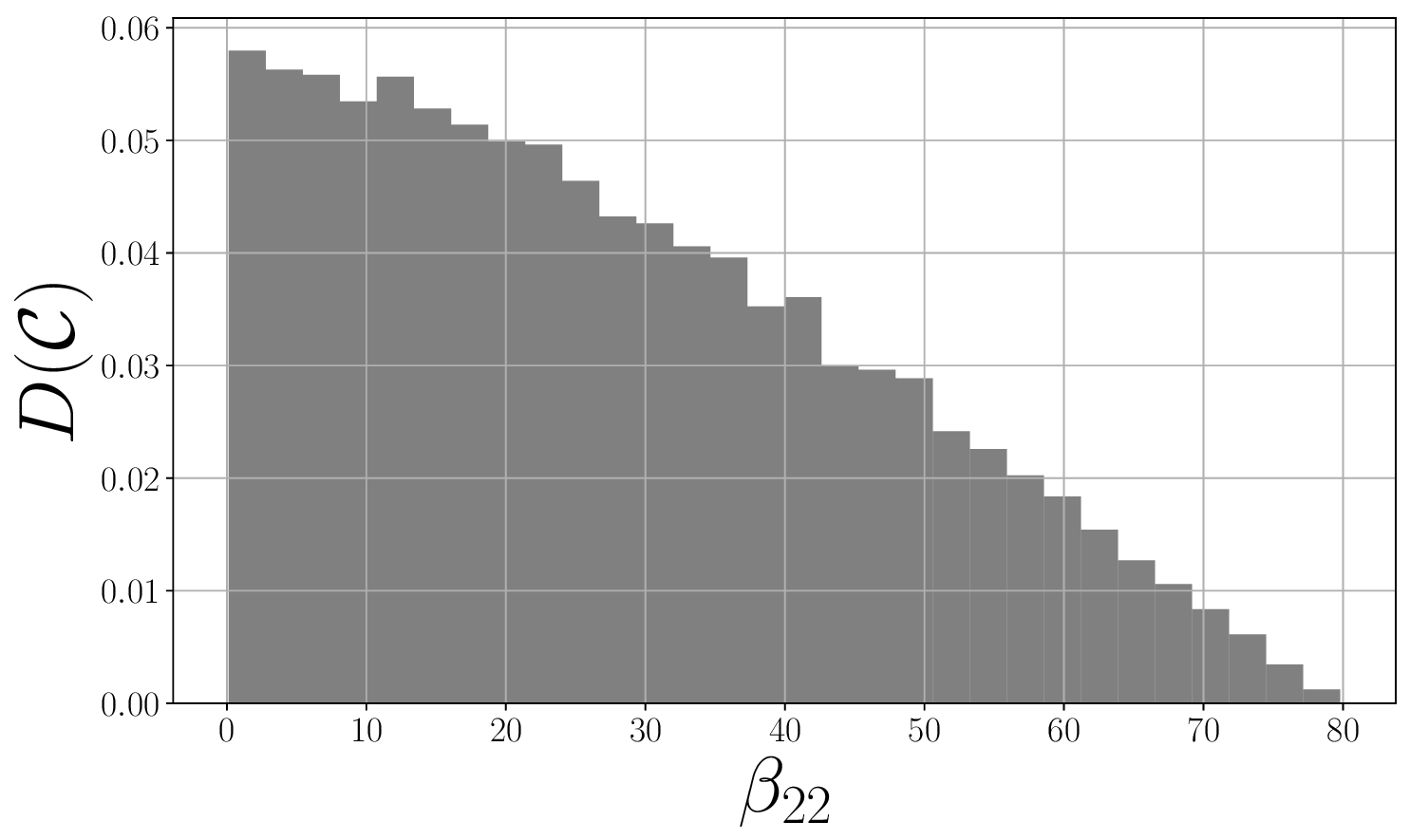}

    \caption{Parameter density distribution ($D$) for the two dimensional search space for Test case 2}
    \label{fig:paramdistfuncT2}
\end{figure}

\begin{figure}[htbp]
    \centering
    \includegraphics[width=0.9\linewidth]{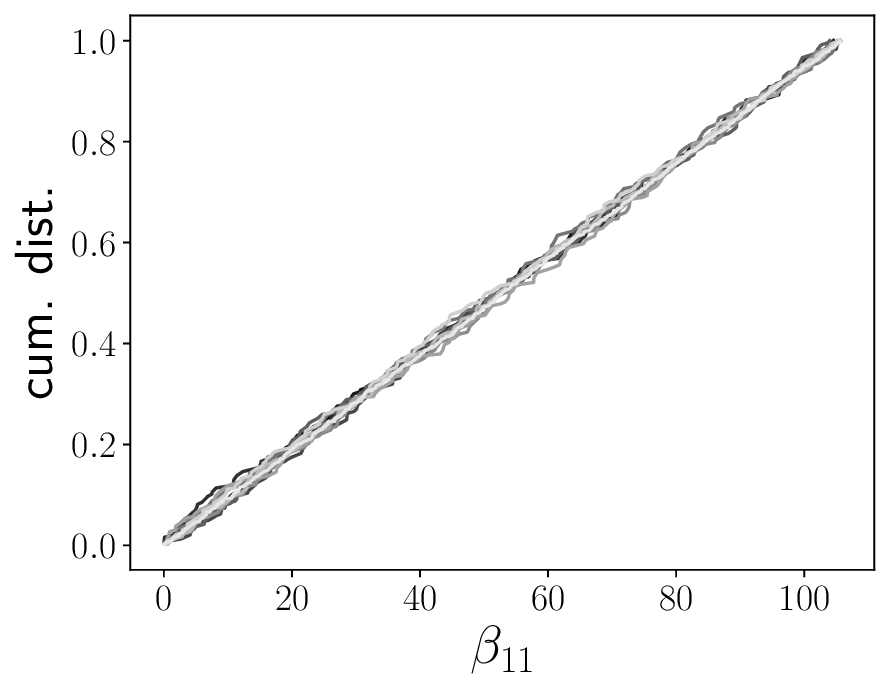}

    \includegraphics[width=0.9\linewidth]{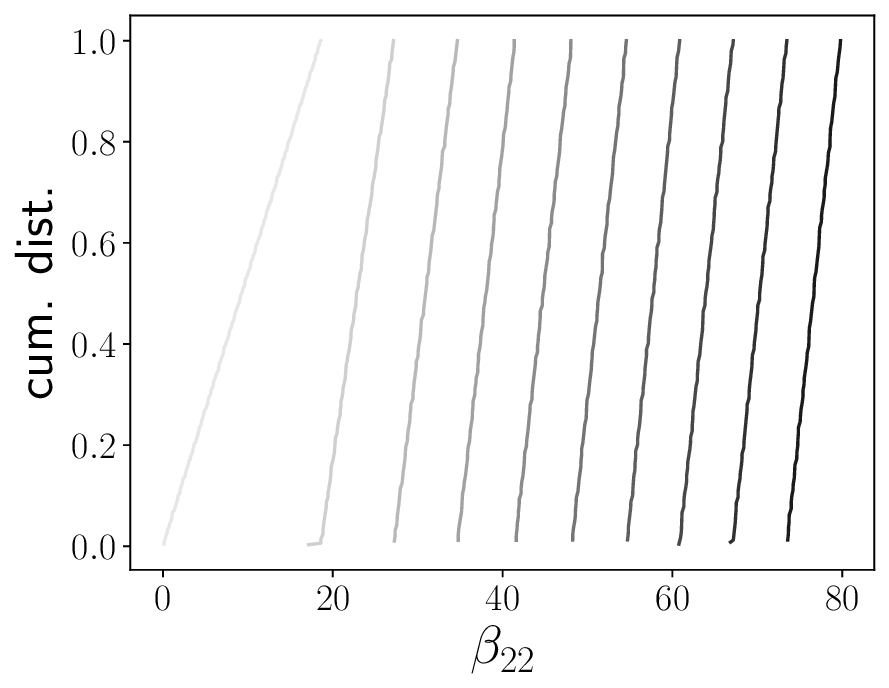}

    \caption{Cumulative distribution function  for the two dimensional search space for Test case 2}
    \label{fig:cumdistfuncT2}
\end{figure}

\begin{figure}[htbp]
    \centering
    \includegraphics[width=0.9\linewidth]{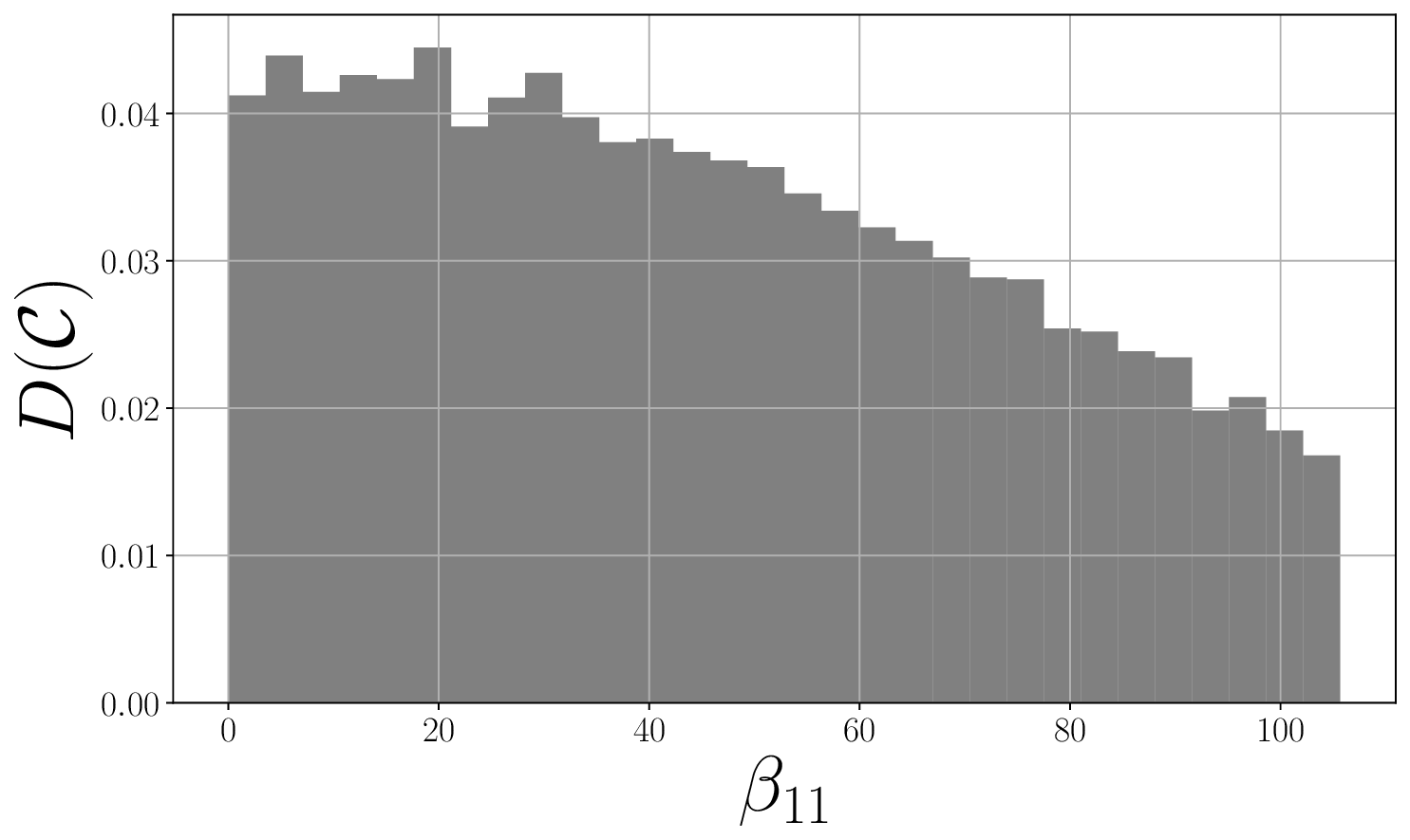}

\includegraphics[width=0.9\linewidth]{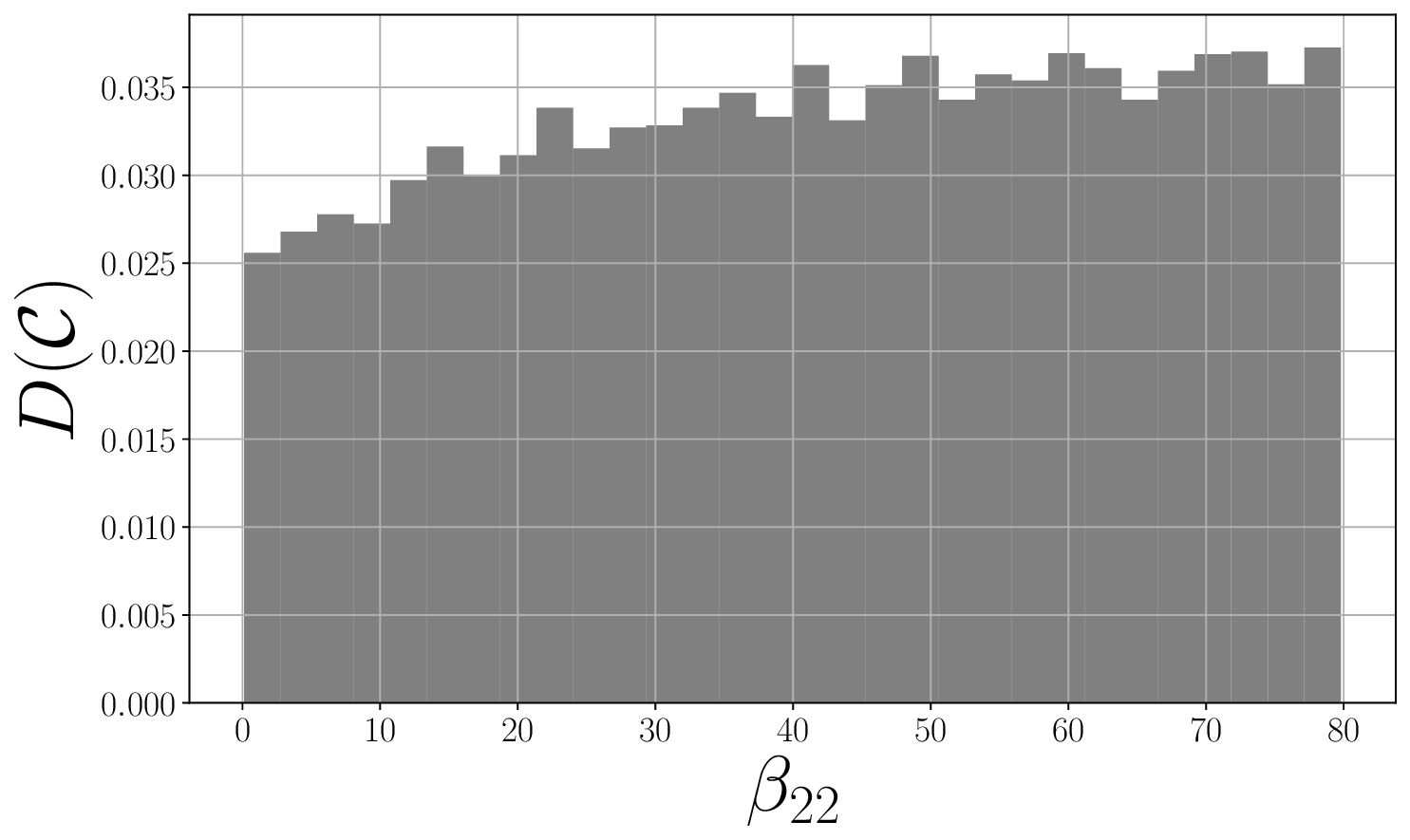}

    \caption{Parameter density distribution ($D$) for the two dimensional search space for Test case 3}
    \label{fig:paramdistfuncT3}
\end{figure}

\begin{figure}[htbp]
    \centering
    \includegraphics[width=0.9\linewidth]{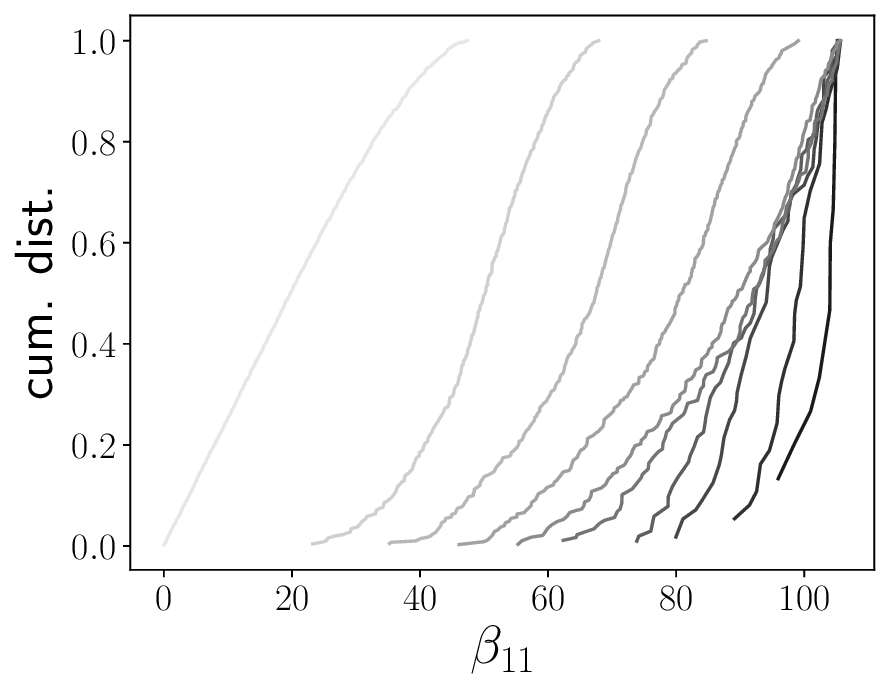}

    \includegraphics[width=0.9\linewidth]{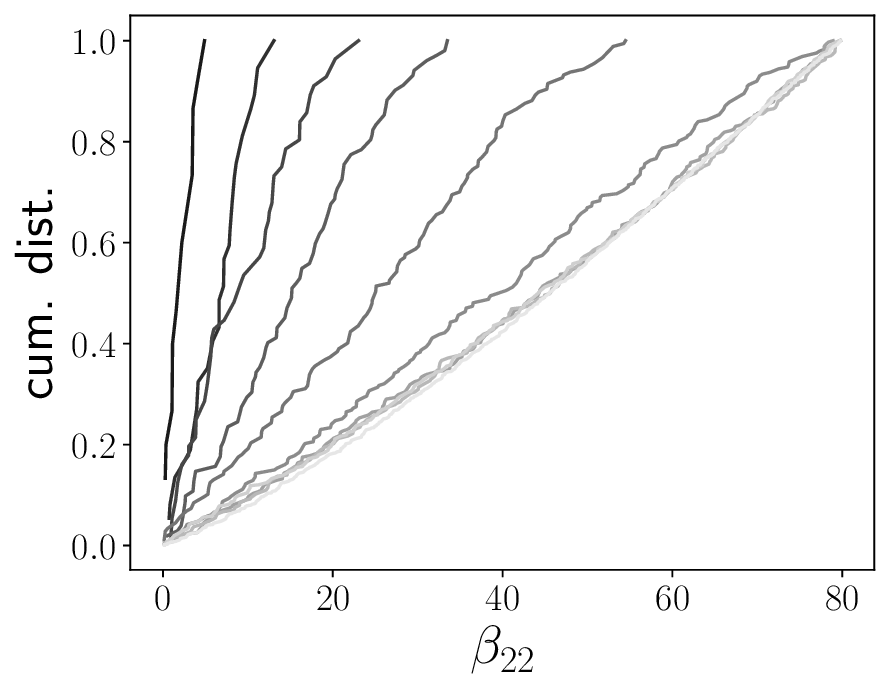}

    \caption{Cumulative distribution function  for the two-dimensional search space for Test case 3}
    \label{fig:cumdistfuncT3}
\end{figure}

We perform a regional sensitivity analysis, as described in Section~\ref{subsec:RSA} for circular inclusions on Test case 1, 2 and 3. In total, the cost function was evaluated for $4096$ samples. We set the threshold $\mathcal{s}=0.12$. For the first and second test case we obtain quite similar results. Thus we start by discussing the results for these two test cases together. The parameter density distributions for Test case 1 and 2 are given in Fig.~\ref{fig:paramdistfuncT1} and Fig.~\ref{fig:paramdistfuncT2}. The parameter density distributions are still close to a uniform distribution for $\beta_{11}$ whereas it seems they have a peak  close to $0$ for $\beta_{22}$. This is an indication, that in the given range of values, $\beta_{22}$ should be chosen small. The cost function is rather insensitive with respect to $\beta_{11}$,  so the choice of this parameter is insignificant. The cumulative distribution function for the friction tensor entries for Test case 1 and 2 are shown in Fig.~\ref{fig:cumdistfuncT1} and Fig.~\ref{fig:cumdistfuncT2} for the parameters sorted into groups with respect to their performance. Lighter coloring  of a graph signifies better performance of the  group corresponding to the graph. The given figures underline the conclusions drawn from the parameter distribution function, indicating that the important parameter in the calibration is $\beta_{22}$, because here the curves for different performing groups  are quite different. 
We now turn to the discussion of the findings for the third test case.
We see that the calibration results here depend on both parameters $\beta_{22}$ and $\beta_{11}$. However, from Fig.~\ref{fig:cumdistfuncT3}, we note that the best performing values for $\beta_{22}$ stem from a large range of values (the light colored curves increase almost linearly). Also, in Fig.~\ref{fig:paramdistfuncT3}, we see that the slope of the histogram for $\beta_{22}$ is less  compared to the slope for $\beta_{22}$ in the other histograms. For $\beta_{11}$ in Fig.~\ref{fig:paramdistfuncT3}, we see that there is a large plateau in the histogram up to values of $40$, that is directly linked to the best performing parameters, as we see by the lightest curve in Fig.~\ref{fig:paramdistfuncT3}. This clearly shows, that the best performing values  for $\beta_{11}$ can be found up to values of about $40$.

We conjecture that the  similarities observed in the RSA results for Test case 1 and 2—specifically the resemblance between Fig.~\ref{fig:paramdistfuncT1} and Fig.~\ref{fig:cumdistfuncT1}, as well as between Fig.~\ref{fig:cumdistfuncT1} and Fig.~\ref{fig:cumdistfuncT2}—are primarily due to the identical shape of the interface $\Sigma$ used in the first two test cases, together with the similar inflow conditions applied.

In contrast, the noticeable differences in Fig.~\ref{fig:paramdistfuncT3} and Fig.~\ref{fig:cumdistfuncT3} compared to their counterparts from Test Cases 1 and 2 indicate that, in general, the calibration is sensitive to both diagonal components of the friction tensor.     

As a conclusion we found, that the chosen test case does have an impact on the optimal  value for the friction tensor. However, there is in general a large interval of reasonable choices for the parameters. This is an indicator, that also for more complex applications (larger simulation domains with complex geometric setup) a calibration of the stress-jump problem is feasible.


\subsubsection{Comparison with an analytical calibration for 1D-flow}\label{subsubsec:comarison1Dflow}

In \citet{Angot_etal_21} a calibration of the stress--jump conditions was performed by comparing exact solutions of the coupled model to solutions of a mathematical model based on the one-domain approach for one-dimensional flow. 
For one-dimensional flow in an isotropic porous medium, the approximation of the first component of the friction tensor  

\begin{equation}
\displaystyle\vec\beta \approx \frac{1}{\sqrt{\mathrm{Da}}}\mathbf{I},
\end{equation}
is derived for a non-dimensionalized equation.
Here, the Darcy number $\mathrm{Da} := \frac{K}{L^2}$ is defined in terms of a characteristic macroscopic length $L$ 
and the scalar-valued permeability $K$. As in \citet{Angot_etal_21}, we choose as representative height the height of the free-flow domain $L=0.5$.
In \citet{Angot_etal_21} the position of the interface in the transition region is also a free parameter.
Since for our case the position of the transition region is not clear, we position the interface as for the other test cases on top of the first row of inclusions.

\begin{figure}[htbp]
    \centering
\includegraphics[width=0.95\linewidth]{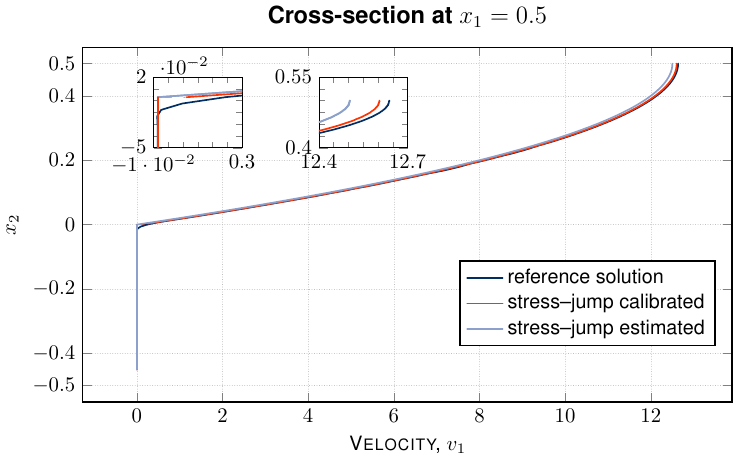}
    \caption{One-dimensional flow, comparison of the first velocity component}
    \label{fig:oneD}
\end{figure}

We perform a calibration for the fourth test case, with the reference solution produced as in Test case 2. The dual annealing optimization was used for a one-dimensional search space. However, the cost function must be adapted to the situation. Because the flow is one-dimensional, only contributions of the first velocity component are considered. 
We define the function
\begin{align}
   \mathcal{D}_g= \frac{|\mathrm g^m-\mathrm g^r|}{\|\mathrm g^r\|_{L _2(\Omega_\FF^t\cup\Omega_\PM^t)}}
\end{align}
on $\Omega_\FF^t\cup\Omega_\PM^t$. 
Here, $\mathrm g^r$ is the pore- and $\mathrm g^m$ the macroscale solution of the quantity $g$ that is either the pressure or the first velocity component in the whole domain. The cost function is defined as
\begin{equation}
\begin{split}
 \begin{alignedat}{1} \label{eq:costfunction1D}
\mathcal{C}(\vec\beta)= &\|\mathcal{D}_{v_1}\|^2_{L_2(\Omega_\FF^t\cup\Omega_\PM^t)}+\|\mathcal{D}_{p}\|^2_{L_2(\Omega_\FF^t\cup\Omega_\PM^t)}
\end{alignedat}
\end{split}
\end{equation}
 This adaptation is necessary because the amplitude of the velocity in the porous medium is significantly smaller compared to that of the free-flow. In the usual formula of the cost function, the contributions from the quantities in the porous medium would thus dominate the value of the cost function in an extreme manner. This is not desirable, since by minimization of the cost function the overall behavior of the system should be improved not only by the behavior in the porous medium.
The result of the calibration is $\vec\beta=3.34$. The value of the friction tensor derived from formula \eqref{eq:IC-beta_1d-SJ} is $70.88$. The respective corresponding values of the relative $L^2$-error are $0.005$ for the calibrated and $0.0147$ for the estimated friction tensor. In Fig.~\ref{fig:oneD} we show the first velocity component for a cross-section at $x=0.5$. 
Close to the interface and on top of the free-flow domain, we observe that the different macroscale models differ. While the numerically calibrated model leads to a large jump in the tangential velocity, the choice of the ad-hoc formula seems to lead to an almost continuous variation on the interface. Since both models lead to small relative errors, the ad-hoc formula might still be a good first choice when employing stress--jump conditions. Furthermore, the results indicate a low sensitivity of the model to the friction tensor for one-dimensional flow, because even though the values for the friction tensor from the numerical Calibration and from the formula are quite different, the resulting simulations are still close. This test case shows, that the stress--jump conditions can also be employed for flow tangential to the interface.

\section{Concluding remarks}\label{chap:conclusion}
For all considered test cases, we were able to calibrate the Stokes--Darcy problem endowed with stress--jump conditions. 
This indicates that the calibrated model aligns well with the reference solution. In particular, it suggests that discrepancies arising from modeling choices—such as the magnitude of the scale‑separation parameter $\epsilon$, the boundary conditions imposed in the porous medium, or the selected size of the REV tend to dominate the behavior of the calibrated model and often outweigh the influence of the interface conditions themselves.
A comparison with the calibrated classical interface conditions showed, that in fact, the calibrated  stress--jump Stokes--Darcy model is not only comparable to the classical model but even performs better for the test cases considered, for all quantities, except the pressure in the free flow. The reasons for the slightly weaker performance of the stress-jump conditions for pressure would require further investigation. Also for tangential flow the stress--jump conditions are applicable and already a simple choice for the friction tensor leads to a well-calibrated model, since the model in this case is quite insensitive to the values of the friction tensor.

For the technical aspects of the calibration we observed that both the dual annealing and the brute force algorithm lead to similar results in most test cases. The most important factor in the calibration is the choice of the search space. Here we see that a good compromise between efficiency and performance for isotropic, orthotropic, and general porous media is already a one-dimensional search space that consists of scalings of the inverted permeability tensor. By this procedure, the complexity of the calibration problem can be reduced significantly. This approach is also a strong indicator supporting the hypothesis used to derive the stress--jump conditions \citep{Angot_etal_17}.

By the regional sensitivity analysis we noted, that the range of admissible values for the friction tensor values is large, rendering the calibration simpler.
These results show, that the stress-jump conditions for flow with arbitrary flow directions are not only applicable but a calibration in more complex situations is promising, because both reduction of the dimension of the search space and insensitivity to the choice of parameters can enhance a simplified cost-efficient calibration.

\backmatter

\bmhead{Acknowledgements}
The work is funded by the Agence nationale de la recherche (ANR, French Research Foundation) ANR-21-CE40-0018.











\bibliography{sn-bibliography}

\begin{appendices}
\section{Experimental convergence analysis}\label{app:EOC}

We implemented a Stokes--Darcy problem with right hand sides in the flow conditions and in the interface conditions. Because of these right hand sides, the Darcy equations have to be supplemented by one additional term stemming from the stabilization term. So we end up with the following coupled problem
\begin{equation}
\begin{split}\label{eq:eqforexactsol}
 \begin{alignedat}{2}
     -\deld\mathbf{T}^\FF\lrp{\vec v^\FF,p^\FF} &= 
    \vec f_\FF
    \, \quad &&\text{in } \ \Omega_\FF \,  , 
    \\
    \nabla \vdot \vec v^\FF &= 0 \quad  &&\text{in } \ \Omega_\FF \,,  \\
         \vec v^\PM &= - \frac{\mathbf{K}}{\mu}\,(\nabla p^\PM -\vec f_\PM) 
         \\&\phantom{=}-\operatorname{curl} \lrp{\mathbf{K}^{-1}\mu\,\vec v^\PM )}
 \, \quad  &&\text{in } \ \Omega_\PM \,  , 
\\
\nabla \vdot \vec v^\PM &= 0 \quad 
&&\text{in } \ \Omega_\PM \,  ,\\
  \vec v^\FF\vdot \vec n&=\vec v^\PM \vdot\vec n \phantom{aaa}\ \ &&\text{on } \ \Sigma\ ,\\
   \llbracket \mathbf{T}({\vec v,p})\vdot \vec n\rrbracket_\Sigma&=\dsp\frac{\mu}{\sqrt{\mathbf{K}}}\vec\beta\vec v^\FF\ +\vec f_\Sigma\ &&\text{on } \ \Sigma \,.
   \end{alignedat}
   \end{split}
\end{equation}
We tested different values of permeability and friction tensors, the results from Section~\ref{subsec:FEMimplementation} were obtained for
the permeability
\begin{align*}
   \mathbf{K}= \begin{pmatrix}
1 &  0.5\\
0.5 & 1
\end{pmatrix},
\end{align*}
and the friction tensor
\begin{align*}
   \vec\beta= \begin{pmatrix}
1 &  0.5\\
0.5 & 2
\end{pmatrix}.
\end{align*}

As exact solution to \eqref{eq:eqforexactsol} we use in the porous medium domain as well as in the free-flow domain the velocity field $\vec v^{ex}=(v_1^{ex},v_2^{ex})$ 
\begin{equation}
\begin{split}\label{eq:exactsolv}
v_1^{ex}&=\sin\Big(\frac{\pi}{2}(x+s_x)\Big)\cos\Big(\frac{\pi}{2}(y+s_y)\Big)\,,\\
v_2^{ex} &= -\sin\Big(\frac{\pi}{2}(y+s_y)\Big)\cos\Big(\frac{\pi}{2}(x+s_x)\Big)\,,
\end{split}
\end{equation}
and the pressure 
\begin{equation}
\begin{split}\label{eq:exactsolp}
p^{ex} &= \cos\Big(\frac{\pi}{2}(x+s_x)\Big)\Big(\sqrt{2}/(2K_{1 1})\\
&\phantom{=}\exp((y+s_y)-0.5) - \sqrt{2}\frac{\pi}{4}\Big)\,.
\end{split}
\end{equation}
For the experimental order of convergence (EOC) presented in Section~\ref{subsec:FEMimplementation} the real parameters $s_x,s_y$ are put as $s_x=1.2$ and $s_y=2.15$.

\section{Parametrization of lune-shaped inclusion}\label{secC}
In this section we provide the parametrization of the lune-shaped inclusion.
The coordinates of the boundary of the inclusion $(x(t), y(t))$ are defined piecewise on $[0,4]$. We have
\begin{align*}
x(t) &= \begin{cases}
         r_1\cos(2\pi(1-t/4)) & \text{if } 0 \leq t \leq 1 \\
         \frac{(r_2-r_1)}{2}\cos(\pi(0.5+t-1)) & \text{if } 1 < t \leq 2 \\
         r_2\cos(2\pi(t/4+0.75)) & \text{if } 2 < t \leq 3 \\
         \frac{r_2-r_1}{2}\cos(\pi(t-3)) \\
         + r_1 + \frac{r_2-r_1}{2} & \text{if } 3 < t \leq 4
       \end{cases}
\end{align*}
for the first and
\begin{align*}
y(t) &= \begin{cases}
         r_1\sin(2\pi(1-t/4)) & \text{if } 0 \leq t \leq 1 \\
         \frac{(r_2-r_1)}{2}\sin(\pi(0.5+t-1))\\-r_1-\frac{(r_2-r_1)}{2} & \text{if } 1 < t \leq 2 \\
         r_2\sin(2\pi(t/4+0.75)) & \text{if } 2 < t \leq 3 \\
         \frac{3(r_2-r_1)}{4}\sin(\pi(t-3)) & \text{if } 3 < t \leq 4
       \end{cases}
\end{align*}
for the second component where $r_1 = 0.2$ and $r_2 = 0.7$.

\section{Box-constraint-algorithm}\label{alg:boxsearch}
We present here for the one-dimensional case the algorithm that determines the box-constraints in preparation to apply the optimization algorithm \ref{alg:boxsearchalg}. 
\begin{algorithm*}
    \caption{1D Adaptive Search Algorithm}\label{alg:boxsearchalg}
    
    \begin{algorithmic}[1]
        \Procedure{AdaptiveSearch1D}{costFunction}
            \State Initialize domain bounds: $lowerBound \gets 0$, $upperBound \gets 100000$
            \State Perform brute force evaluation
            \State Create grid points $P = \{p_1, p_2, ..., p_{60}\}$ uniformly spaced in $[lowerBound, upperBound]$
            
            \While{iteration < maxIterations}
                \For{i = 1 to 60}
                    \State Evaluate cost function: $value_i \gets costFunction(p_i)$
                \EndFor
                
                \State Find optimal value: $optValue \gets \min(value_1, value_2, ..., value_{60})$
                \State Set threshold: $\theta \gets optValue + 0.02$
                
                \State Initialize lower bound tracking
                \State $lowerBoundFound \gets False$
                \State $newLowerBound \gets None$
                
                \For{i = 1 to 59}
                    \If{NOT $lowerBoundFound$ AND $value_i \leq \theta$ AND $value_{i+1} > \theta$}
                        \State $newLowerBound \gets p_{i+1}$
                        \State $lowerBoundFound \gets True$
                    \EndIf
                \EndFor
                
                \If{$lowerBoundFound$}
                    \State $lowerBound \gets newLowerBound$
                \EndIf
                
                \State Initialize upper bound tracking
                \State $upperBoundFound \gets False$
                \State $newUpperBound \gets None$
                
                \For{i = 59 downto 2}
                    \If{NOT $upperBoundFound$ AND $value_i > \theta$ AND $value_{i-1} \leq \theta$}
                        \State $newUpperBound \gets p_i$
                        \State $upperBoundFound \gets True$
                    \EndIf
                \EndFor
                
                \If{$upperBoundFound$}
                    \State $upperBound \gets newUpperBound$
                \EndIf
                
                \State iteration $\gets$ iteration + 1
            \EndWhile
            
            \Return $(lowerBound, upperBound)$
        \EndProcedure
    \end{algorithmic}
\end{algorithm*}





\end{appendices}



\end{document}